\documentclass[amsmath,amssymb,aps, physrev]{revtex4-2}
\usepackage{color}
\usepackage{graphicx}% Include figure files
\usepackage{dcolumn}% Align table columns on decimal point
\usepackage{bm}% bold math
\usepackage{amssymb}
\usepackage{amsmath}
\usepackage{tabularx}
\usepackage{tikz}
\usepackage[utf8]{inputenc}
\usepackage[T1]{fontenc}
\usepackage{mathptmx}
\usepackage{etoolbox}
\usepackage[normalem]{ulem}
\usepackage{array}
\usepackage{booktabs}           % MAKE TABLE BETTER 
\usepackage{caption}            % CAPTIONS OF FIGURES
\usepackage{subcaption}         % CAPTIONS OF SUBFIGURES

\newcolumntype{C}{>{\centering\arraybackslash}X} 

\begin{document}

\title{\textbf{Primary and secondary motions in an annular plane Couette flow}}% 

\author{Rémi Macadré}
\affiliation{Institut de M\'ecanique des Fluides de Toulouse (IMFT), CNRS, Universit\'e de Toulouse, France}
\affiliation{%
Laboratoire de G\'enie Chimique (LGC), CNRS-INPT-UPS, Universit\'e de Toulouse, France}
\affiliation{TotalEnergies S.E., P\^ole d'\'etudes et de Recherche de Lacq (PERL), France}
\affiliation{FR FERMAT, Universit\'e de Toulouse, CNRS, Toulouse, France}

\author{F. Risso}%
\affiliation{Institut de M\'ecanique des Fluides de Toulouse (IMFT), CNRS, Universit\'e de Toulouse, France}
\affiliation{FR FERMAT, Universit\'e de Toulouse, CNRS, Toulouse, France}

\author{O. Masbernat}
\affiliation{%
Laboratoire de G\'enie Chimique (LGC), CNRS-INPT-UPS, Universi\'e de Toulouse, France%\\This line break forced% with \\
}
\affiliation{FR FERMAT, Universit\'e de Toulouse, CNRS, Toulouse, France}%

\author{R. Belt}
\affiliation{TotalEnergies S.E., P\^ole d'\'etudes et de Recherche de Lacq (PERL), France}

\date{\today}

\begin{abstract}
Axisymmetric direct numerical simulations are carried out to study the hydrodynamics of a laminar, stationary, incompressible, Newtonian, single-phase flow in an Annular Plane Couette (APC) channel. This configuration is that of a Straight Plane Couette (SPC) flow but curved around itself to form an annulus. These simulations are validated by Particle Image Velocimetry measurements at different Reynolds numbers. The flow is analyzed using three dimensionless parameters: the channel aspect ratio $A_c$, which controls the effects of sidewall confinement, the channel curvature ratio $C_r$, which affects the centrifugal forces due to curvature, and the Reynolds number ${\rm Re}$. The rotation of the top annular plate generates a main flow in the azimuthal direction, while generating a secondary recirculation flow in the plane of the channel cross-section, due to the presence of a centrifugal force difference. As a result, the vertical profile of the azimuthal velocity deviates from the classical linear profile of the SPC flow, adopting an unexpected S-shape, similar to that observed in a turbulent plane Couette flow. Depending on the values of $A_c$ and $C_r$, the flow exhibits a wide range of behaviors, from a quasi-2D flow with a homogeneous shear rate at moderate ${\rm Re}$ with appropriate geometrical parameters ($A_c \gtrsim 5$, $C_r \lesssim 0.1$), to a complex 3D flow otherwise. Whereas it is laminar, the APC flow shares a strong analogy with a turbulent Taylor Couette (TC) flow. At large Reynolds numbers, the velocity gradients concentrate near the wall and the flow reaches an asymptotic regime where the torque scales as ${\rm Re}^\alpha$ and the flow structure becomes independent of ${\rm Re}$. Such properties make the APC flow an interesting configuration for fundamental investigations as a complement of the TC flow when shear and gravity are required to be in the same plane. In particular, it can be used to investigate the rheology of a dispersed two-phase mixture similarly to the work done by Yi et al. \cite{YI2021,Yi2022} in the TC device.
\end{abstract}

\maketitle

\section{\label{sec:sec1}Introduction}

Studies in fluid mechanics often use configurations of reference to investigate relevant physical processes. In particular among these reference cases, Couette-type and Poiseuille flows find a wide range of applications. They are widely used to study transitions to turbulence \cite{AVILA2023, ORLANDI2015, TILLMARK1992}, turbulence \cite{ECKHARDT2000, Grossmann2016, TELBANY1980, PIROZZOLI2011}, rheology \cite{YI2021,ABBAS2017, LEIGHTON86, Yi2022} or transitions of flow configurations \cite{ANGELI98, ANGELI2000, POUPLIN2011} in multiphase flows.  

Couette-type devices can be classified into two distinct geometries: (1) the Taylor-Couette (TC) flow, characterized by two concentric rotating cylinders, and (2) the Straight Plane Couette flow (SPC), characterized by one static wall and one moving wall. The fluid sandwiched between these walls in both geometries is propelled forward by the shear caused by the wall motion, leading to classify these configurations as shear-driven flows. Poiseuille flows, on the other hand, result from a pressure difference along a cylindrical or rectangular pipe. These flows are referred to as pressure-driven flows.  

While Poiseuille flows are predominant in industrial applications, Taylor-Couette flows are more commonly used for research purposes. This preference stems from the ability to conduct long-term observations thanks to their periodicity, as well as maintaining a uniform shear rate. While the SPC flow keeps the advantage of a constant controllable shear rate, it has no periodicity due to inlet and outlet boundary conditions. However, in multiphase flow studies, it is desirable to have a plane of shear comprising the vertical direction, as in the SPC flow, rather than perpendicular to gravity, as in the TC flow, to study the effects of both gravity and shear on the vertical migration of the phases. A configuration that combines both the vertical plane of shear and flow periodicity to allow for long-term observations is the Annular Plane Couette (APC) flow. This geometry is basically that of a SPC flow that would be curved around itself to form a closed loop.

The APC flow has often been used for the study of geophysical flows \cite{FUKUDA1980, STAUDT2017, BAAR2020, CHARRU2004}. It is well-known that due to channel curvature, secondary flows of Prandtl's first kind \cite{CHIN2020} (generated by centrifugal forces) develop in the cross-section, resulting in a complex 3D flow. Previous studies of the hydrodynamics in this geometry mainly focused on the turbulent regime \cite{BOOIJ2003, GHARABAGHI2007, POPE2006} and on the optimal rotation speed ratio between top and bottom walls, in order to minimize the curvature effect on the radial shear-stress distribution at the bottom wall \cite{BOOIJ1999, BOOIJ1994, PETERSEN2012,YANG2015473}. However, the hydrodynamics of this flow remains poorly addressed in the laminar regime. This issue is tackled in the present work for single-phase flows, in view of considering this configuration as a reference case for the study of complex multiphase flows, as already done to characterize the rheology of suspensions \cite{GUAZZELLI2018}. In this work, the flow in the APC is investigated in the laminar regime, with direct numerical simulations using OpenFOAM, and further validated by Particle Image Velocimetry (PIV) in an experimental APC device at various Reynolds numbers.

The present paper is organized as follows: in Section~\ref{sec:sec2}, the geometry of the APC flow is detailed and the non-dimensional parameters characteristic of the flow hydrodynamics in laminar regime are introduced. Section \ref{sec:sec3} describes the numerical method and the experimental set-up used for the simulations and the measurements respectively. The results obtained and their physical interpretation are discussed in Section~\ref{sec:sec4}. Then, scaling laws for both the torque exerted by the fluid on the moving wall and the secondary flow are presented in section~\ref{sec:sec4}. Finally, section~\ref{sec:sec6} summarizes the main results.

\section{\label{sec:sec2}Annular Plane Couette flow}

\subsection{Description of the geometry}

The Straight Plane Couette (SPC) flow refers to the standard plane Couette configuration, distinguished by a top moving wall that drives the fluid in the longitudinal direction through shearing in an unbound channel. Its laminar flow is characterized by a vertical linearity of the longitudinal velocity profile, resulting in a constant shear rate $\dot{\gamma}_m = U_{top}/H$, with $U_{top}$ the velocity of the top moving wall and $H$ the channel height. The laminar SPC flow has been shown to be stable to infinitely small disturbances at any Reynolds number \cite{ROMANOV1973}, in contrast to what happens in the APC flow (Section~\ref{sec:sec4}.B.1).

\begin{figure}[h!]
     \centering
     \begin{subfigure}[h]{0.48\textwidth}
         \centering
         \includegraphics[width=0.8\textwidth]{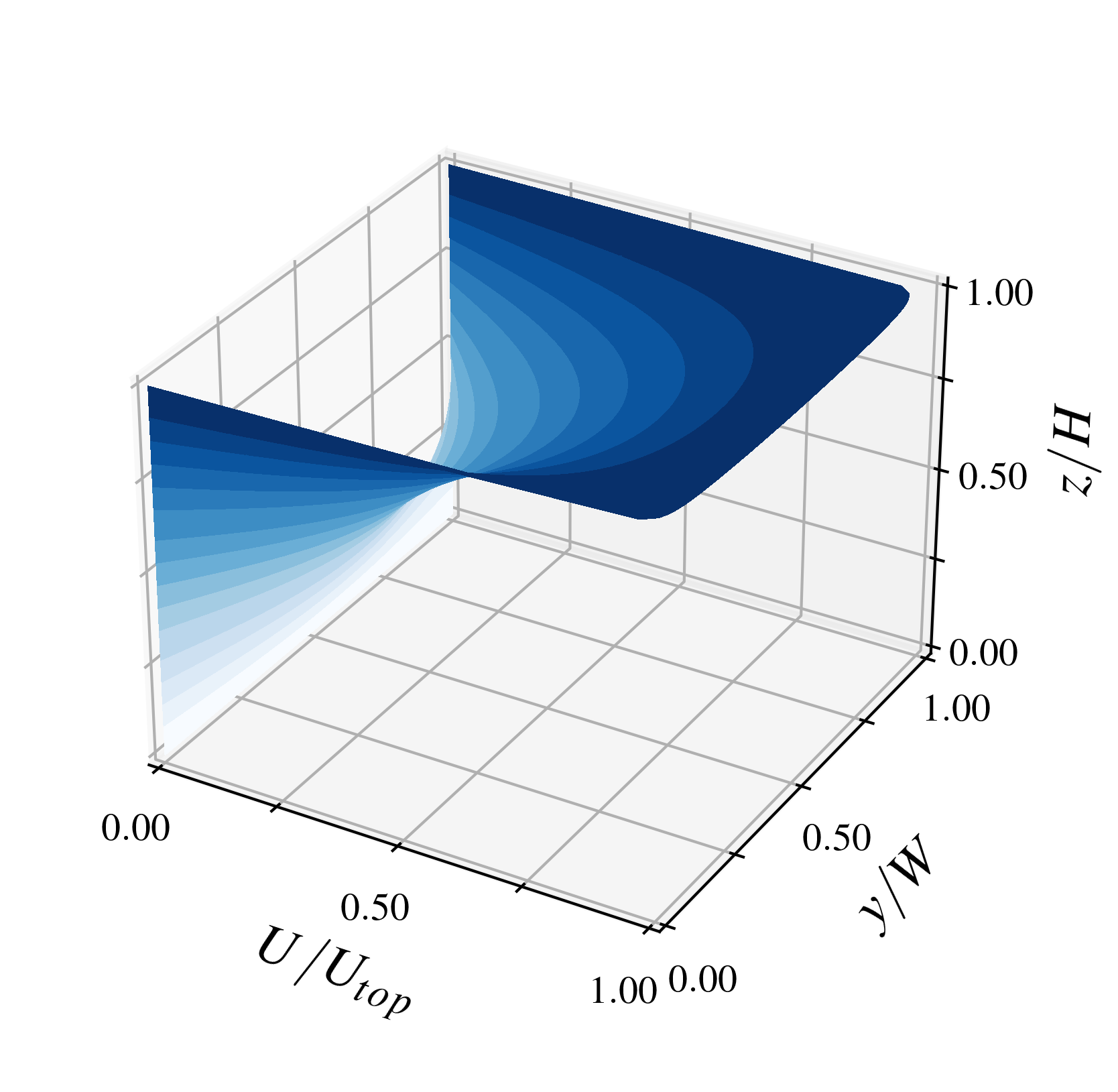}
         \caption{$A_c =  1$}
         \label{fig:Af_1}
     \end{subfigure}
     \hfill
     \begin{subfigure}[h]{0.48\textwidth}
         \centering
         \includegraphics[width=0.8\textwidth]{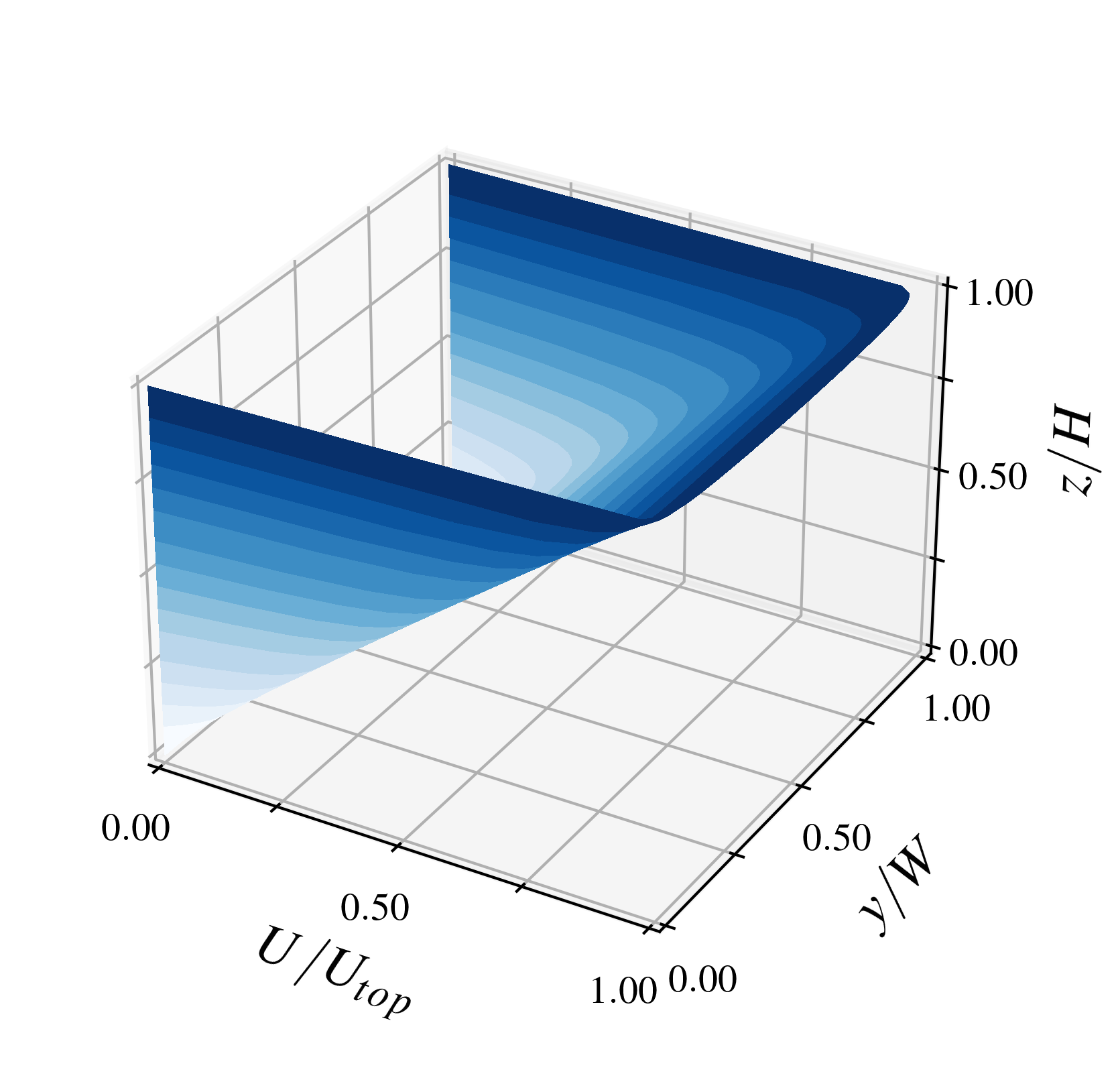}
         \caption{$A_c = 10$}
         \label{fig:Af_10}
     \end{subfigure}
     \caption{Velocity profile in a Straight Confined Plane Couette (SCPC) flow for different aspect ratios $A_c = W/H$.}
     \label{fig:SCPC}
\end{figure}

Adding lateral walls to the straight channel shapes the configuration into a rectangular-section channel, confining the fluid inside and turning the SPC flow into the Straight Confined Plane Couette (SCPC) flow. The analytical solution for the SCPC flow is also well-established \cite{WIKICOUETTE} and depends on the channel aspect ratio $A_c = W/H$, with $W$ the channel width. 
The influence of the channel aspect ratio is illustrated in Fig.~\ref{fig:SCPC}. When $A_c = 1$ (Fig.~\ref{fig:Af_1}), the channel has a square cross-section and significant wall effects distort the velocity field. In contrast, when $A_c = 10$ (Fig.~\ref{fig:Af_10}), the solution reverts to the classical linear profile of the SPC flow in the majority of the channel, but the flow remains confined at the side-wall regions due to the influence of the no-slip boundary conditions.

\begin{figure}[h!]
\centering
\includegraphics[scale=1.3]{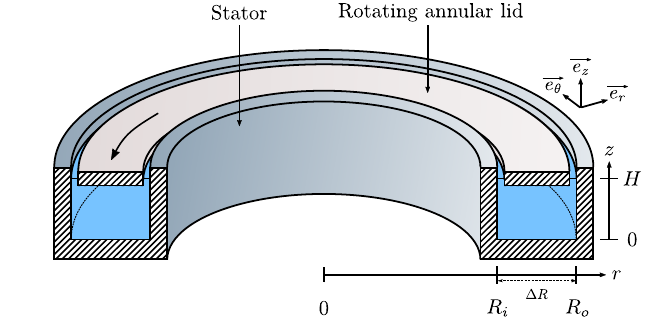}
\caption{\label{fig:apc_geom} Schematic of the Annular Plane Couette (APC) flow. The fluid is driven by the rotation of the top annular plate.}
\end{figure}

Finally, a configuration capable of accommodating long-term observations and the benefits of the SPC flow can be designed. A well-suited geometry that fulfill these requirements is the configuration termed here as Annular Plane Couette (APC) flow, which is schematized in Fig.~\ref{fig:apc_geom}. The channel is shaped as an annular ring of rectangular cross-section, with a height $H$, inner radius $R_i$ and outer radius $R_o$, corresponding to a channel width of $\Delta R = R_o - R_i$. The middle radius is defined by $R_{\text{mid}} = (R_o + R_i)/2$ and the middle height is noted $Z_{\text{mid}} = H/2$. The cylindrical coordinate system ($r, \theta, z$) is used for the analysis. By rotating the top annular plate, the fluid within the channel is propelled forward, generating a primary flow in the azimuthal direction. As in the SCPC configuration, the confinement due to the lateral walls influences the velocity field. Additionally, the curvature of the channel introduces a centrifugal force difference along the radius that further distorts the velocity field, as it will be discussed in section \ref{sec:sec4}. 

\subsection{Dimensionless parameters}

The analysis of the hydrodynamics of the laminar, stationary, axisymmetric, incompressible, Newtonian, single-phase flow within the APC channel is conducted using three different dimensionless parameters: the channel Reynolds number ${\rm Re}$, the channel aspect ratio $A_c = \Delta R / H$ and the channel curvature ratio $C_r = \Delta R / R_o$.

The Reynolds number ${\rm Re}$ is based on the channel's half-height $H/2$ and its half maximum velocity $U_{top,max}/2 = \Omega R_o /2$ at the top. It writes
\begin{equation}
    \label{eq:rec}
    {\rm Re} = \dfrac{\rho \Omega R_o H}{4 \mu},
\end{equation}
where $\Omega$ is the rotation velocity of the top annular plate, $\rho$ the fluid density and $\mu$ its dynamic viscosity. 

A criterion for the transition to turbulence was obtained experimentally by Tillmark \textit{et al.} \cite{TILLMARK1992} in a SPC flow. They found that the onset of the turbulence occurs at ${\rm Re} \sim 360$, while the transition to fully developed turbulence takes place around ${\rm Re} \sim 500$. However, since the SPC flow is linearly stable at any Reynolds number, that experimental result probably depends on uncontrolled finite disturbances and cannot be considered as universal. We ignore whether a critical Reynolds number beyond which the APC flow becomes unstable exists. Steady laminar solutions of APC numerical simulations for ${\rm Re} \gtrsim 360$ are thus not irrelevant, but have to be considered with care regarding their application to real flows at large ${\rm Re}$.

The geometrical effects are twofold. As in the SCPC configuration, the channel aspect ratio number $A_c = \Delta R / H$ influences the velocity field due to the sidewall confinement. In addition, the effect of the channel curvature on the velocity field is analyzed by considering the channel curvature ratio $C_r = \Delta R/R_o$. This ratio can be derived by comparing the centrifugal stress difference across the radius $\tau_{c} = \rho \Omega^2 R_o \Delta R$ and the viscous stress $\tau_{\mu} = \mu \Omega R_o / H$. This results in
the product of the Reynolds number and the curvature ratio as
\begin{equation}
    \dfrac{\tau_{c}}{\tau_{\mu}} = \dfrac{\rho \Omega R_o H}{\mu} \dfrac{\Delta R}{R_o} = 4 Re C_r.
\end{equation}

\noindent
While $\tau_{c}/\tau_{\mu}$ characterizes the effects of the centrifugal force difference on the flow, we separate the analysis by varying independently the channel Reynolds number (defined in Eq.~\ref{eq:rec}) and $C_r$, since the results are expected to depend on both $\rm{Re}$ and $C_r$.

\section{\label{sec:sec3}Description of numerical simulations and experimental set-up}

\subsection{Methods and flow conditions}

Direct numerical simulations of the APC flow are performed to better understand the relevance of such a configuration as a reference flow. These simulations are done with the open-source CFD software \textit{OpenFOAM}. The Navier-Stokes equations are solved by the \textit{simpleFOAM} module, which uses the finite-volume method and the SIMPLE algorithm for velocity-pressure coupling. The equations are discretized using Gauss linear numerical schemes of the second order accuracy in space. A steady, incompressible and Newtonian single-phase flow is assumed here.

\begin{figure}[h!]
\centering
%\begin{tikzpicture}
\includegraphics[width=0.432\textwidth]{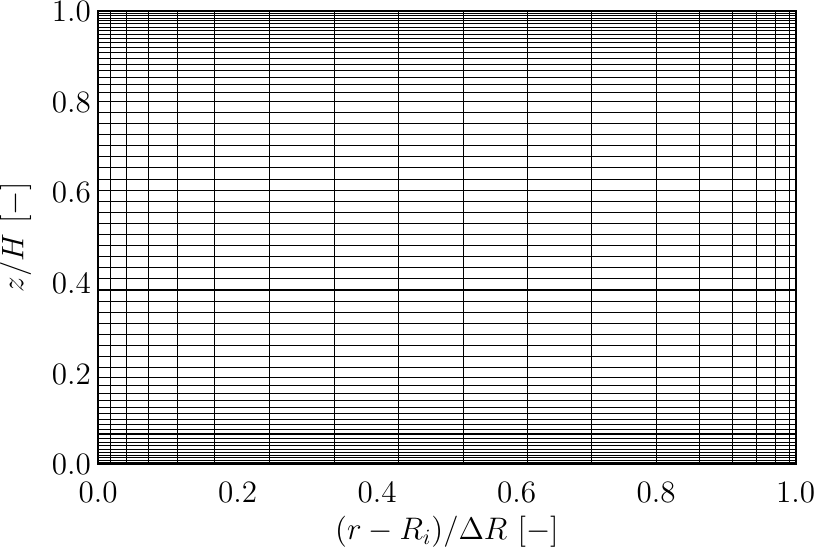}
%\node[text width=2cm] at (7.0,-0.7) {$\Uw = \Omega r$};
%\end{tikzpicture}
\caption{\label{fig:bc_mesh} Example of the numerical mesh in the radial-vertical plane ($r-z$).}
\end{figure}

Since the flow is axisymmetric, only the radial-vertical section of the channel is simulated to ensure faster computations. The mesh is exclusively composed of hexahedron cells. Regular meshing is applied to the central region of the section while the mesh is refined near the walls to provide more accurate results of the boundary layer in both vertical and radial directions. An example of the numerical mesh in the radial-vertical plane is shown in Fig.~\ref{fig:bc_mesh}.

Regarding the boundary conditions, a no-slip condition is enforced on every wall for the velocity. The top annular plate rotates at a fixed angular velocity $\Omega$, causing the azimuthal velocity of the fluid at the top wall to vary radially as $U_{top}(r) = \Omega r$ while the other walls remain stationary. The zero-gradient boundary condition is used for pressure at each wall. Finally, periodic boundary conditions are applied in the azimuthal direction for the velocity and pressure.

Various simulations are performed by varying the inner radius $R_i$, the height $H$ of the channel and the rotational speed $\Omega$ of the top annular plate, leading to the exploration of the range of dimensionless parameters summarized in Table~\ref{tab:table1}.

\begin{table}[h!]
    \captionsetup{width=\textwidth}
    \centering
    \begin{tabular}{>{\centering\arraybackslash}m{3cm} >{\centering\arraybackslash}m{3cm} >{\centering\arraybackslash}m{3cm} >{\centering\arraybackslash}m{3cm}}  
        \toprule
        Parameters & $\rm{Re}$ & $A_c$ & $C_r$\\
        \midrule
        Ranges & 5 to 500 & 1 to 10 & 0.1 to 0.9\\
        \bottomrule
    \end{tabular}
    \caption{\label{tab:table1}Ranges of dimensionless parameters for the numerical simulations.}
\end{table}

\subsection{Mesh validation}

To validate the numerical mesh, the most critical case corresponding to the highest ${\rm Re}$, $C_r$ and the lowest $A_c$ is examined. The selected parameters for this case are ${\rm Re} = 500$, $C_r = 0.9$ and $A_c =~1$. Different mesh sizes are compared for the radial-vertical plane: $N_r \times N_z = 70 \times 250,\text{ } 140 \times 500$ and $ 280 \times 1000$. Regarding mesh convergence, the vertical and radial profiles of two key quantities are considered for validation. The first quantity is the dimensionless azimuthal velocity $U_{\theta}/\Omega r_{\text{val}}$, where $ r_{\text{val}}$ is the middle radius $R_{\text{mid}} $ for vertical profiles and the outer radius $R_o$ for radial profiles. The second quantity is the dimensionless shear rate $\dot{\gamma}/\dot{\gamma}_m$, with $ \dot{\gamma} = \sqrt{\left (2 \uuline{S} : \uuline{S} \right )}$ based on the strain-rate tensor $\uuline{S}$, and $\dot{\gamma}_m = \Omega R_{\text{mid}}/H$. With this definition, $\dot{\gamma}/\dot{\gamma}_m$ is equal to unity in a SPC flow and constitutes a relevant scale for the mesh assessment in an APC flow since it accounts for all velocity-gradient components.

\begin{figure}[h!]
     \centering
     \begin{subfigure}[h]{0.48\textwidth}
         \centering
         \includegraphics[width=\textwidth]{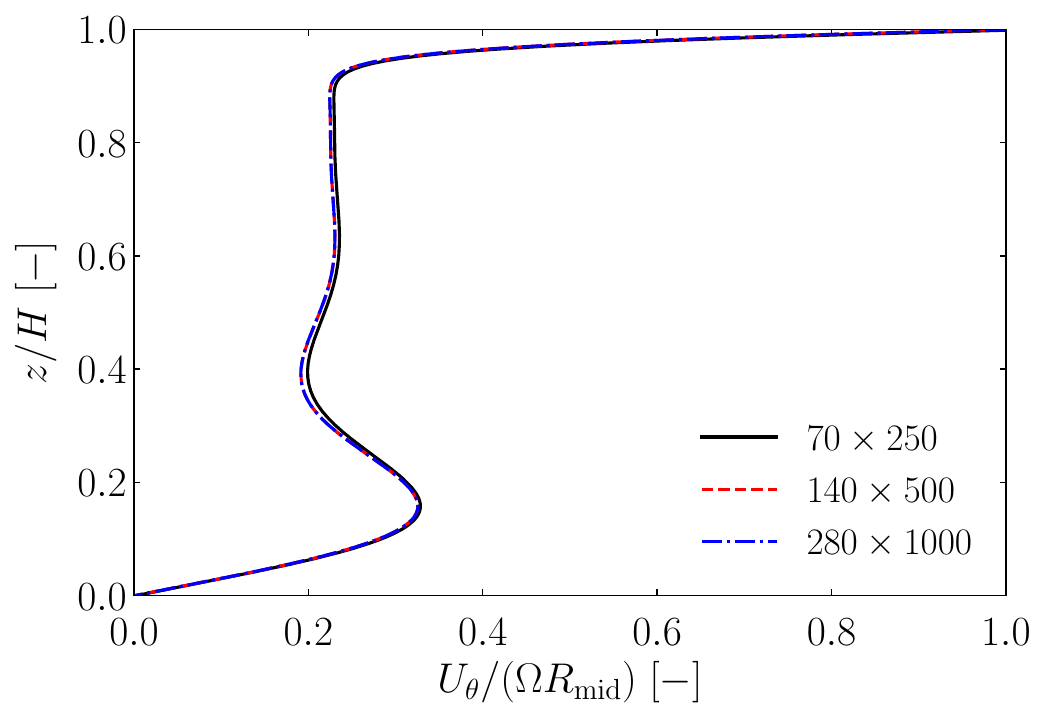}
         \caption{Profile of $U_{\theta}(z)$ at $r=R_{\text{mid}}$.}
         \label{fig:vali_ut_z}
     \end{subfigure}
     \hfill
     \begin{subfigure}[h]{0.48\textwidth}
         \centering
         \includegraphics[width=\textwidth]{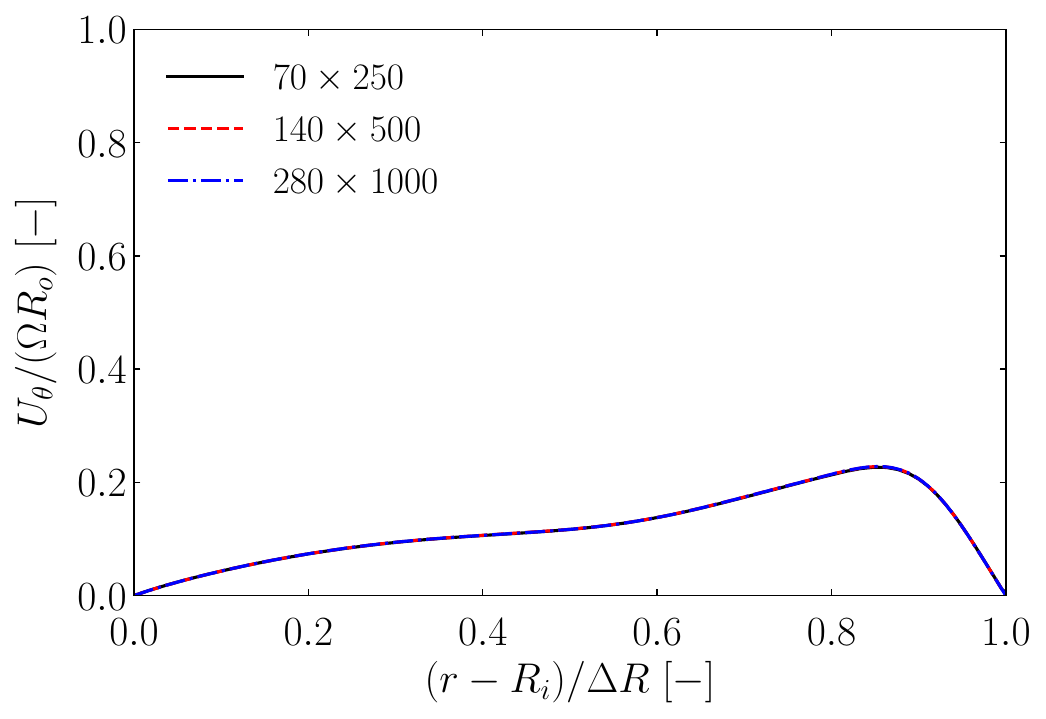}
         \caption{Profile of $U_{\theta}(r)$ at $z=Z_{\text{mid}}$.}
         \label{fig:vali_ut_r}
     \end{subfigure}
     \caption{Azimuthal velocity profiles in the most critical case (${\rm Re} = 500$, $A_c = 1$, $C_r = 0.9$) for mesh validation. The results of three different mesh sizes are presented: $N_r \times N_z = 70 \times 250$, $140 \times 500$ and $280 \times 1000$.}
     \label{fig:ut_validation}
\end{figure}

Fig.~\ref{fig:ut_validation} shows the profiles of the azimuthal velocity $U_{\theta}$ along the vertical direction at the middle radius $r = R_{\text{mid}}$ and in the radial direction at the middle height $z = Z_{\text{mid}}$. The profiles corresponding to the three different meshes exhibit similarity. Convergence is achieved even at the lowest resolution ($N_r \times N_z = 70 \times 250$), with only a small difference in the vertical profiles in the lower central region.

\begin{figure}
     \centering
     \begin{subfigure}[h]{0.48\textwidth}
         \centering
         \includegraphics[width=\textwidth]{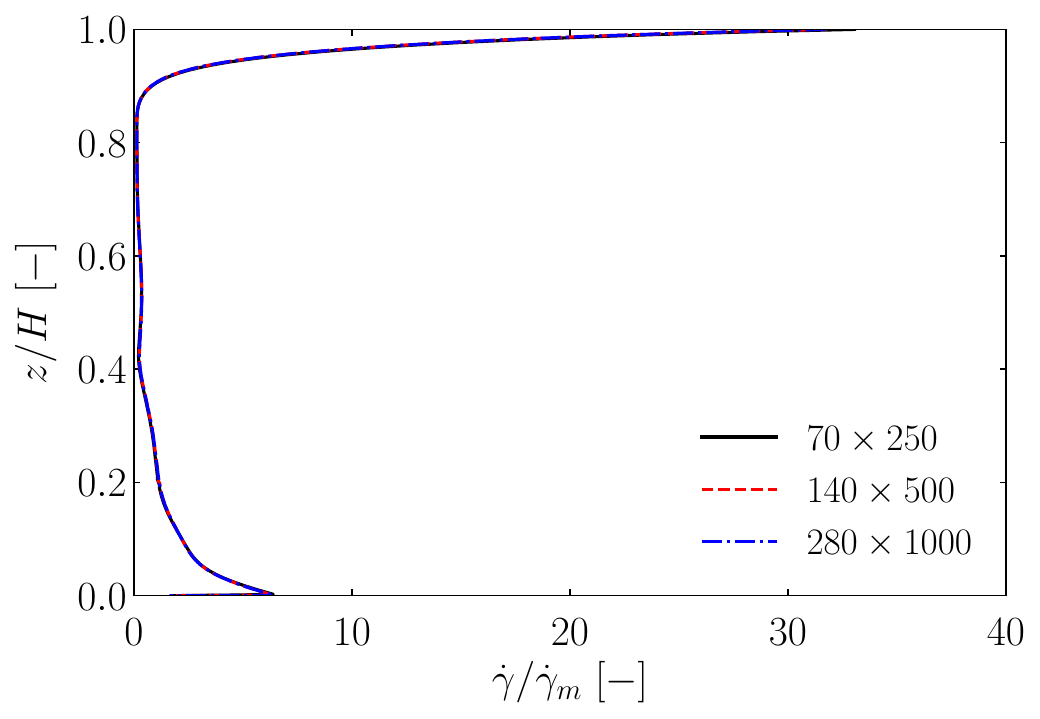}
         \caption{Profile of $\dot{\gamma}(z)$ at $r=R_{\text{mid}}$.}
         \label{fig:vali_gd_z}
     \end{subfigure}
     \hfill
     \begin{subfigure}[h]{0.475\textwidth}
         \centering
         \includegraphics[width=\textwidth]{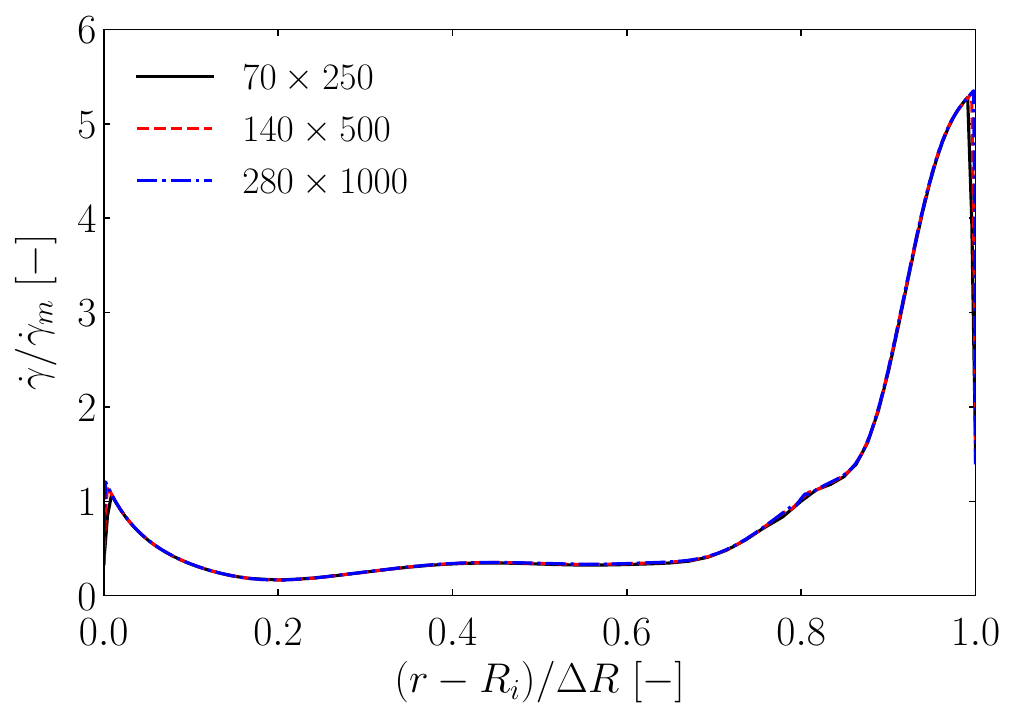}
         \caption{Profile of $\dot{\gamma}(r)$ at $z=Z_{\text{mid}}$.}
         \label{fig:vali_gd_r}
     \end{subfigure}
     \caption{Shear-rate profiles in the most critical case (${\rm Re} = 500$, $A_c = 1$, $C_r = 0.9$) for mesh validation. 3 different mesh sizes are plotted: $70 \times 250$, $140 \times 500$ and $280 \times 1000$.}
     \label{fig:gd_validation}
\end{figure}

Fig.~\ref{fig:gd_validation} displays the shear-rate profiles for the same conditions as in Fig.~\ref{fig:ut_validation}. Once more, the simulations present only minute differences across all mesh resolutions. However, for the radial profiles, the boundary layers at the outer and inner radii exhibit steep shear-rate gradients, and increasing the mesh resolution improves the capture of this effect.

In conclusion, the results show that all tested meshes are suitable. To properly capture the gradients of shear rate at the boundary layers, the resolution of the $140 \times 500$ mesh grid emerges as the most pertinent choice for this study. Subsequently, to keep a same mesh resolution for each value of $A_c$ or $C_r$, the number of cells in both radial ($N_r$) and vertical ($N_z$) directions will be adjusted to keep the same cell sizes, $\Delta r$ and $\Delta z$, as in the $140 \times 500$ case.

\subsection{Experimental set-up}\label{sec:expsetup}

\begin{figure}
\centering
\includegraphics[width=0.48\textwidth]{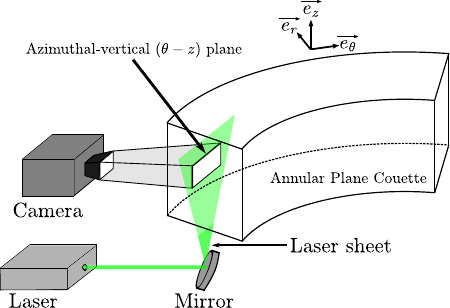}
\caption{\label{fig:exp} Experimental set-up for the PIV measurement of the azimuthal velocity.}
\end{figure}

To validate the numerical simulations of the APC flow, experiments are carried out using the set-up illustrated in Fig.~\ref{fig:exp}. The geometrical dimensions of the channel are $R_i = 0.48 $~m and $R_o = 0.58$~m, corresponding to a curvature ratio $C_r = 0.17$. The height of the channel is adjustable through the top plate vertical position and is set at $H = 0.01$~m, thus yielding an aspect ratio of $A_c = 10$. The channel is made of transparent PMMA to allow for optical measurements. The flow field of a low viscosity oil (n-Dodecane, CAS: 112-40-3, $\rho = 749$ kg$\cdot$m$^{-3}$, $\mu = 1.34 \times 10^{-3}$ Pa$\cdot$s) is measured using Particle Image Velocimetry (PIV) with a PIV system that uses a double pulsed Nd:YAG laser (532~nm, 2~×~120~mJ, Nanopiv – Litron Lasers). The liquid is seeded using rhodamine-doped polystyrene particles of average diameter $d_{\text{mean}} = 10$ \textmu m (PS-FluoRed from \textit{microParticles}, Germany). The azimuthal-vertical plane at middle radius $r = R_{\text{mid}}$ is illuminated by a laser sheet (Fig.\ref{fig:exp}) to obtain a vertical profile of the azimuthal velocity $U_{\theta}$. Recording of instantaneous flow images is achieved using an sCMOS camera (Imager sCMOS from \textit{LaVision}, Germany) with a definition of 5.5 million pixels and a high-pass filter. Image pairs are captured at a rate of 9 Hz, and for each case, 1000 instantaneous velocity fields are recorded. The images are then processed by DaVis software (\textit{LaVision}, Germany), using a decreasing interrogation window size (from $64 \times 64$ pixels$^2$ to $32 \times 32$ pixels$^2$) with 50 $\%$ overlap. Standard cross-correlation with Fast Fourier Transform is employed to determine the corresponding spatially-averaged displacement vectors. The spatial resolution of the velocity fields is $600$ \textmu m. The uncertainties in the PIV measurements are obtained by using the method of Wieneke \cite{WIENEKE2015}.

\section{\label{sec:sec4}Results and discussion}

This section outlines the results of the numerical simulations obtained with meshes of same resolution, $\Delta r$ and $\Delta z$, as the $140\times500$ mesh grid previously validated. The hydrodynamic characteristics are initially scrutinized on a representative case, followed by a parametric study within the ranges of dimensionless parameters given in Table~\ref{tab:table1}.

\subsection{Study of a representative case}

\begin{figure}[h!]
     \centering
     \begin{subfigure}[h]{0.48\textwidth}
         \centering
         \includegraphics[width=\textwidth]{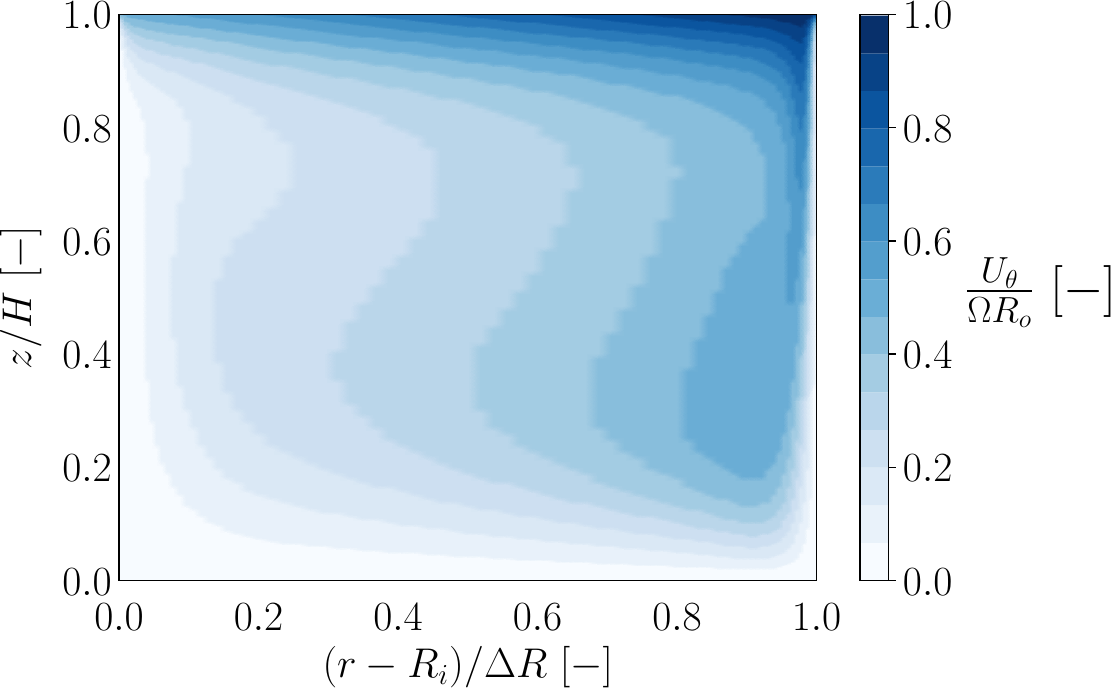}
         \caption{Primary flow ($U_{\theta}$).}
         \label{fig:snap_ut}
     \end{subfigure}
     \hfill
     \begin{subfigure}[h]{0.48\textwidth}
         \centering
         \includegraphics[width=\textwidth]{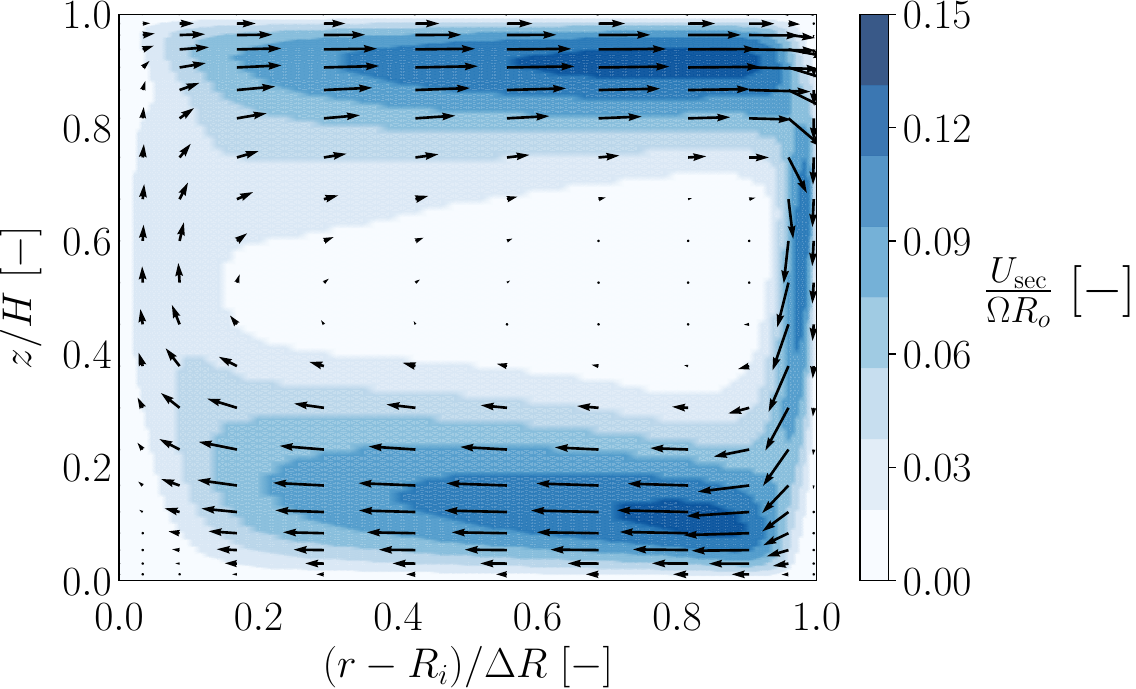}
         \caption{Secondary-flow $(U_r,U_z)$.}
         \label{fig:snap_used}
     \end{subfigure}
     \caption{Velocity fields of the primary and secondary flows in the radial-vertical plane from direct numerical simulations at ${\rm Re} = 280$, $C_r = 0.5$ and $A_c = 5$. For the secondary flow, the color map represents its magnitude $U_{\text{sec}} = \sqrt{U_r^2 + U_z^2}$ and the vector field its velocity vectors.}
     \label{fig:snap_alone}
\end{figure}

Here, the analysis focuses on a representative case characterized by ${\rm Re} = 280$, $C_r = 0.5$ and $A_c = 5$, corresponding to intermediate values of the dimensionless parameters. The rotation of the top annular plate shears the fluid within the channel, creating a flow in the azimuthal direction. This flow is referred to as primary flow (Fig.~\ref{fig:snap_ut}). Simultaneously, the curvature of the channel induces a centrifugal force difference along the radius that generates a secondary flow in the cross-section. Indeed, this centrifugal force propels the fluid in the top region towards the outer radius, pushing the fluid in the outer region downward, due to the presence of the lateral walls. Consequently, the fluid in the bottom region moves towards the inner radius, and due to mass conservation, the fluid in the inner region ascends, completing the formation of a radial-vertical recirculation cell. That flow, involving radial and vertical velocity components, is referred to as secondary flow (Fig.~\ref{fig:snap_used}). One interesting feature is that, in the inner region, the secondary flow spreads upwards over a larger area than in the three other near-wall regions. This results in a lower intensity in the inner region.

The corresponding flow fields are depicted in Fig.~\ref{fig:snap_alone}. For the azimuthal velocity (Fig.~\ref{fig:snap_ut}), the flow exhibits a high vertical gradient at the bottom and top regions, deviating from the constant shear-rate of the SPC flow. This is caused by the presence of the secondary flow in these regions, which disturbs the linear velocity profile. Moreover, its influence is also visible in the outer region where the secondary flow goes downward. Regarding the secondary flow (Fig.~\ref{fig:snap_used}) made of vertical and radial velocities, it is mainly concentrated near the wall regions with a magnitude up to $15 \%$ of the primary-flow magnitude, resulting in a significant influence on the primary flow.

\begin{figure}
\centering
\includegraphics[width=0.432\textwidth]{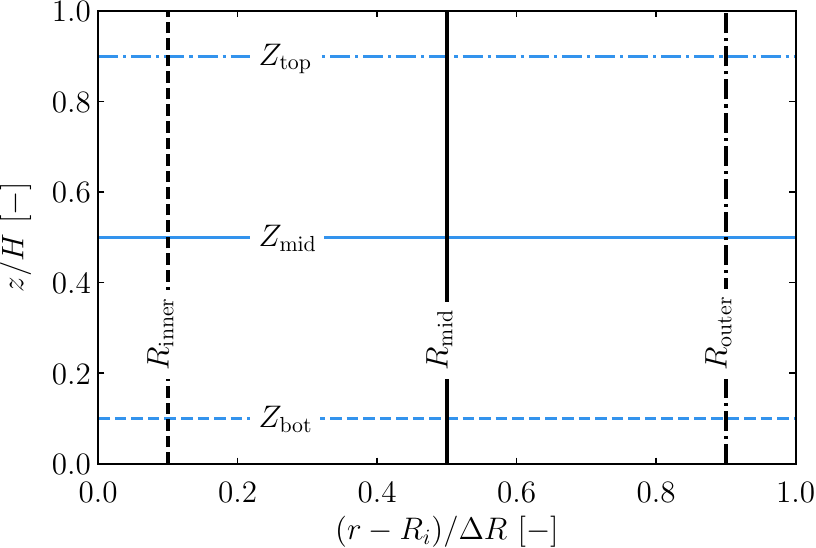}
\caption{\label{fig:profile_scheme} Vertical and radial positions of the profiles in the radial-vertical plane. The colors and line patterns correspond to their respective positions in Figs~\ref{fig:ut_alone} and \ref{fig:gd_alone}. For the radial profiles: $Z_{\text{top}}$ (\textcolor{blue}{-$\cdot$-}), $Z_{\text{mid}}$ (\textcolor{blue}{-}), $Z_{\text{bot}}$ (\textcolor{blue}{- -}). For the vertical profiles: $R_{\text{outer}}$ (-$\cdot$-), $R_{\text{mid}}$ (-), $R_{\text{inner}}$ (- -).}
\end{figure}

\begin{figure}
\label{fig:ut_alone_z_all}
     \centering
     \begin{subfigure}[h]{0.48\textwidth}
         \centering
         \includegraphics[width=\textwidth]{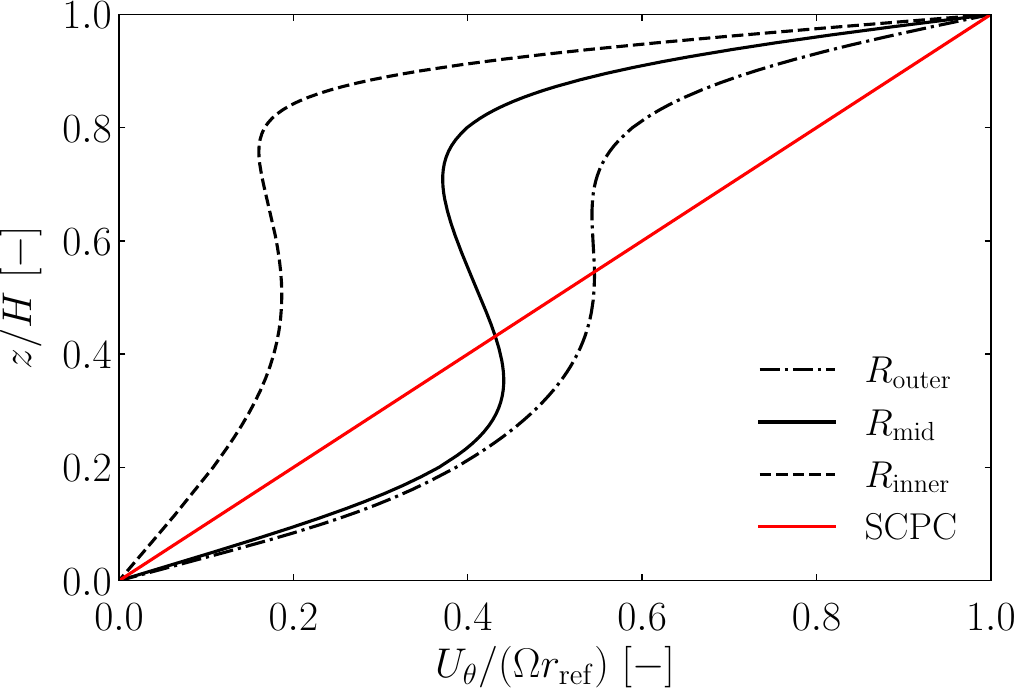}
         \caption{Profile of $U_{\theta}(z)$ at different radii.}
         \label{fig:ut_alone_z}
     \end{subfigure}
     \hfill
     \begin{subfigure}[h]{0.48\textwidth}
         \centering
         \includegraphics[width=\textwidth]{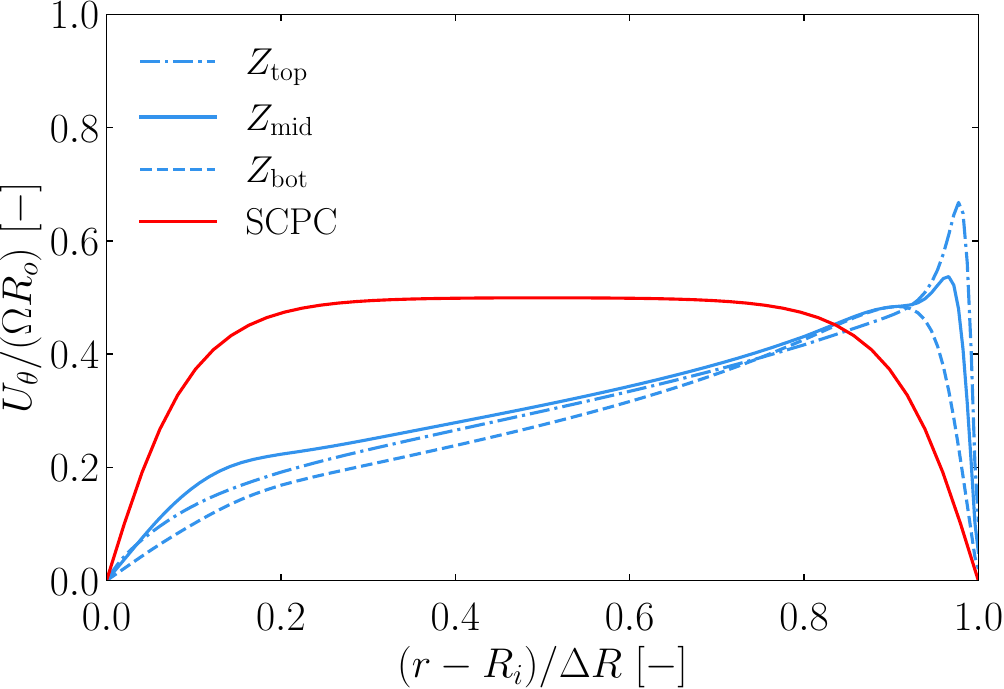}
         \caption{Profile of $U_{\theta}(r)$ at different heights.}
         \label{fig:ut_alone_r}
     \end{subfigure}
     \caption{Azimuthal velocity profiles obtained in the simulations at ${\rm Re} = 280$, $C_r = 0.5$ and $A_c = 5$. The profile positions are given in Fig.~\ref{fig:profile_scheme}. The red lines correspond to the analytical solution of the Straight Confined Plane Couette (SCPC) flow taken at the middle radius $R_{\text{mid}}$ for the vertical profiles and middle height $Z_{\text{mid}}$ for the radial profiles.}
     \label{fig:ut_alone}
\end{figure}

Profiles of the azimuthal velocity $U_{\theta}$ and shear rate $\dot{\gamma}$ at various radial ($R_{\text{inner}}$, $R_{\text{mid}}$, $R_{\text{outer}}$) and vertical positions ($Z_{\text{bot}}$, $Z_{\text{mid}}$, $Z_{\text{bot}}$), defined in Fig.~\ref{fig:ut_alone}, are now presented. The vertical profiles of the azimuthal velocity (Fig.~\ref{fig:ut_alone_z}) are normalized by the top-plate tangential velocity at the considered radial position $\Omega r_{\text{ref}}$, where $r_{\text{ref}} = R_{\text{inner}}$, $R_{\text{mid}}$ or $ R_{\text{outer}}$. They deviate from the typical linear profile associated with the laminar SCPC flow (corresponding to the red curve of Fig.~\ref{fig:ut_alone_z}). Unexpectedly, they exhibit a closer resemblance to the S-shaped profile, characteristic of a turbulent SPC flow, with the presence of significant gradients near the walls. Meanwhile, the profiles in the central region remain almost flat (low vertical gradient), and this pattern holds for each considered radius. The influence of the sidewall confinement being already accounted for in the SCPC channel, this distortion may thus be ascribed to the effect of the centrifugal force difference resulting from the curvature of the channel. Additionally, the gap between each plateau in the central region of the velocity does not appear to be proportional to $r_{\text{ref}}$ and diminishes as $r_{\text{ref}}$ increases. This suggests that the secondary flow does not distort the profiles in the same way at each radius. Finally, the near-wall regions exhibit distinct linear slopes, especially at the top wall, while the curves for the middle and outer radii coincide at the bottom wall, indicating a variation of the intensity of the secondary flow in these regions.

As for the vertical profiles, the radial profiles of the azimuthal velocity (Fig.~\ref{fig:ut_alone_r}) undergo significant changes in the APC flow compared to a SCPC flow. The profile in the central region is distorted into a quasi-linear profile with a slope that does not depend on the vertical position in the channel. However, differences are visible in the near-wall regions. At the outer radius, all profiles drop to zero within almost the same boundary layer thickness, but they peak at different values depending on the vertical position of the considered profile.

\begin{figure}
     \centering
     \begin{subfigure}[h]{0.48\textwidth}
         \centering
         \includegraphics[width=\textwidth]{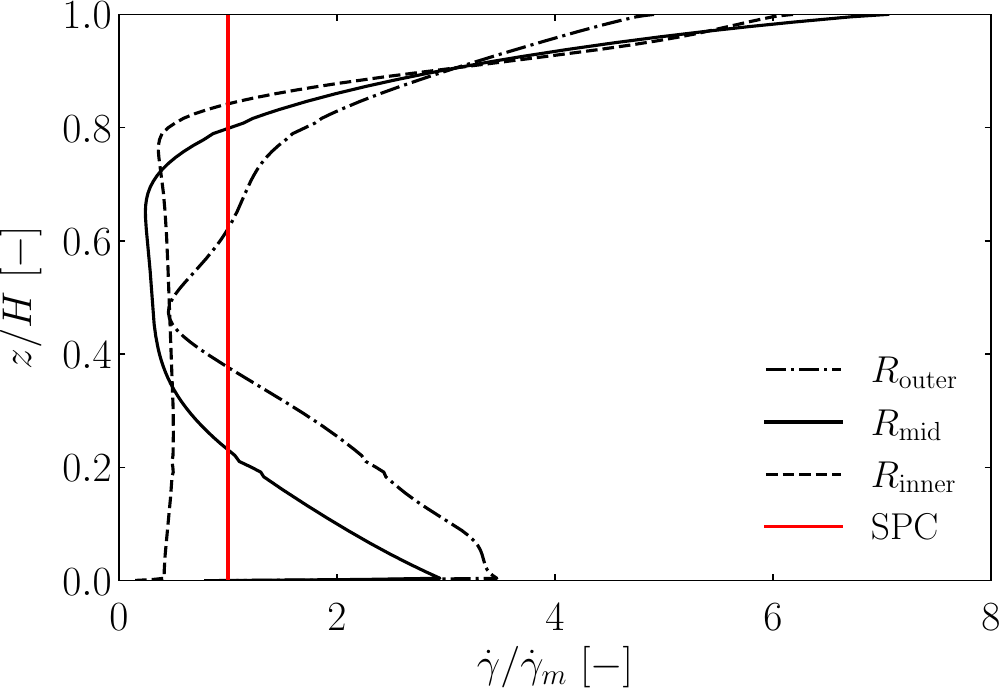}
         \caption{Profile of $\dot{\gamma}(z)$ at different radii.}
         \label{fig:gd_alone_z}
     \end{subfigure}
     \hfill
     \begin{subfigure}[h]{0.48\textwidth}
         \centering
         \includegraphics[width=\textwidth]{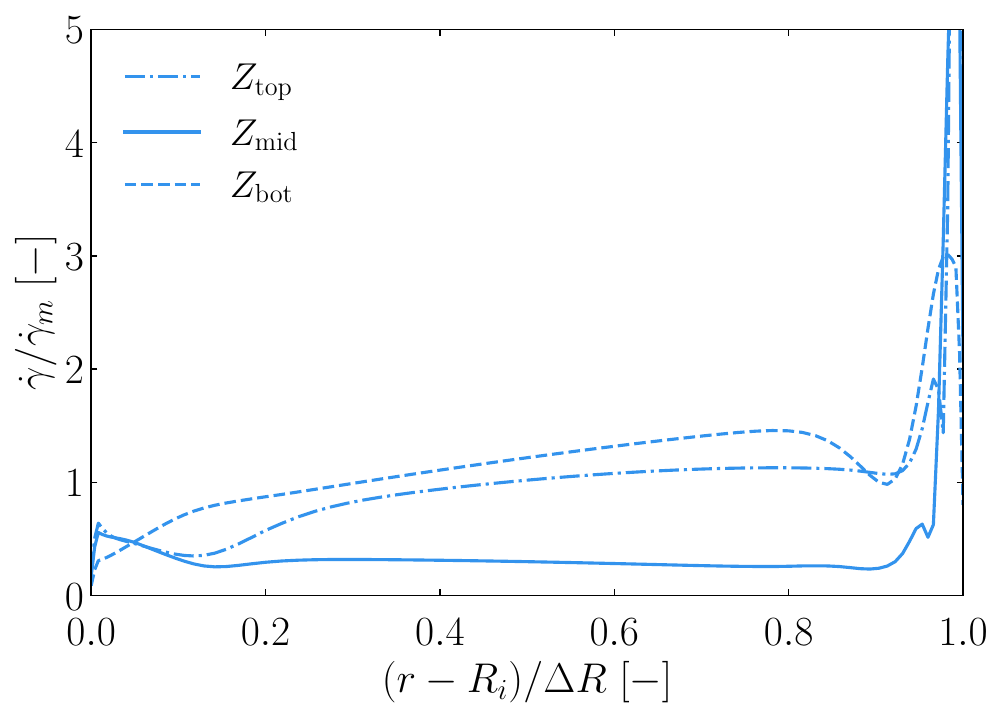}
         \caption{Profile of $\dot{\gamma}(r)$ at  different heights.}
         \label{fig:gd_alone_r}
     \end{subfigure}
     \caption{Computed Shear-rate profiles for ${\rm Re} = 280$, $C_r = 0.5$ and $A_c = 5$. The profile positions are given in Fig.~\ref{fig:profile_scheme}. The red line correspond to the shear rate of the Straight Plane Couette (SPC) flow.}
     \label{fig:gd_alone}
\end{figure}

The shear-rate profiles, obtained in the same conditions as in Fig.~\ref{fig:ut_alone}, are reported in Fig.~\ref{fig:gd_alone}. Since $ \dot{\gamma}$ accounts for local velocity gradients in all directions, it is influenced by the presence of gradients in both the primary and secondary flows. An examination of the vertical profiles (Fig.~\ref{fig:gd_alone_z}) reveals several key features. Specifically, at $r_{\text{ref}}=R_{\text{inner}}$ and $R_{\text{mid}}$, the shear rate in the central region is constant and consistently lower than the value in a SPC flow (red curve in Fig.~\ref{fig:gd_alone_z}). This is consistent with the absence of a secondary flow (Fig.~\ref{fig:snap_used}) and the flatness of the vertical profiles of the azimuthal velocity in the central region (Fig.~\ref{fig:ut_alone_z}). This is not the case for shear rate profile at $r_{\text{ref}}=R_{\text{outer}}$, where the secondary flow is strong with stiff gradients in the wall region. Near the top and bottom regions, at $r_{\text{ref}}=R_{\text{mid}}$ and $R_{\text{outer}}$, the shear rate exhibits important gradients issued from both primary and secondary flows. At $r_{\text{ref}}=R_{\text{inner}}$ the profile remains flat in the bottom region, due to the low magnitude of the secondary-flow gradients.

Regarding the radial profiles of the shear rates (Fig.~\ref{fig:gd_alone_r}), the central part of the profile at middle height $Z_{\text{mid}}$ stays notably low and flat, as a result of the small primary and secondary-flow gradients in this region. The bottom and top-height profiles show a slight increase of the shear rate along the radial coordinate, from the inner to the outer wall. In the vicinity of the outer wall, the shear rate abruptly grows due to the magnitude of secondary-flow gradients in this region.

In conclusion, the velocity vertical profile of the primary flow exhibits an unexpected S-shape as in a turbulent plane Couette flow. The presence of a secondary-flow is evidenced at the bottom, top, and outer wall zones, forming a recirculating loop in the channel section. This secondary flow is impacting the primary velocity field within a boundary layer thickness near the wall. In these regions, significant gradients of both primary and secondary flow are developing, considerably increasing the local shear rate. However, this general flow structure can be modulated by adjusting the channel Reynolds number ${\rm Re}$ and geometrical parameters $A_c$ and $C_r$. In the next section, the influence of these flow parameters is explored and discussed.

\subsection{Parametric study}

In this section, we present the results obtained from simulations using the ranges of dimensionless parameters given in Table.~\ref{tab:table1}. Only the vertical profiles at the middle radius $R_{\text{mid}}$ and the radial profiles at the half-height $Z_{\text{mid}}$ of the azimuthal velocity are focused on.

\subsubsection{Influence of the Reynolds number}

\begin{figure}
     \centering
     \begin{subfigure}[h]{0.48\textwidth}
         \centering
         \includegraphics[width=\textwidth]{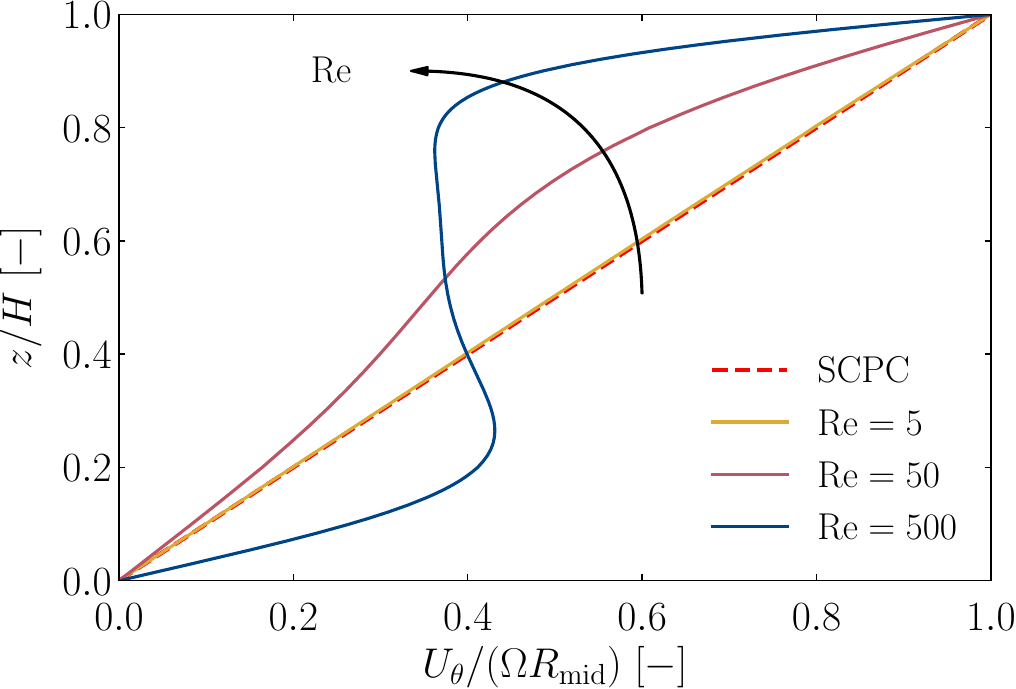}
         \caption{Vertical profiles of the azimuthal velocity $U_{\theta}(z)$ at $r=R_{\text{mid}}$.}
         \label{fig:utz_re}
     \end{subfigure}
     \hfill
     \begin{subfigure}[h]{0.48\textwidth}
         \centering
         \includegraphics[width=\textwidth]{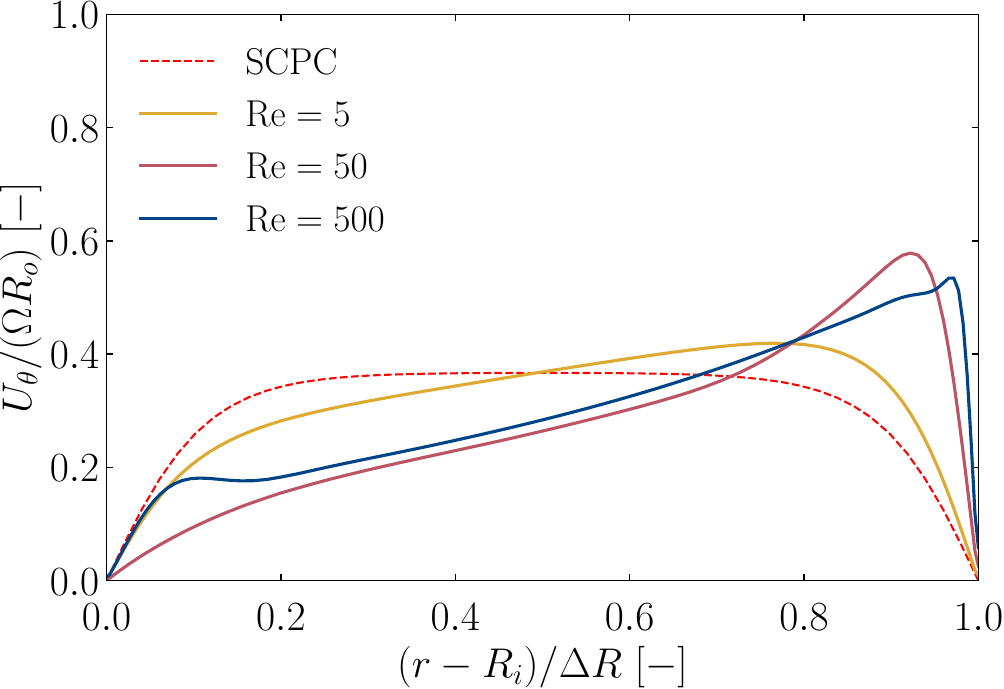}
         \caption{Radial profiles of the azimuthal velocity $U_{\theta}(r)$ at $z=Z_{\text{mid}}$.}
         \label{fig:utr_re}
     \end{subfigure}
     \caption{Azimuthal velocity profiles for $A_c = 5$ and $C_r = 0.5$ at different Reynolds number. The red dotted lines represent the corresponding profiles of the SCPC flows.}
     \label{fig:ut_re}
\end{figure}

\begin{figure}
\centering
\includegraphics[width=0.48\textwidth]{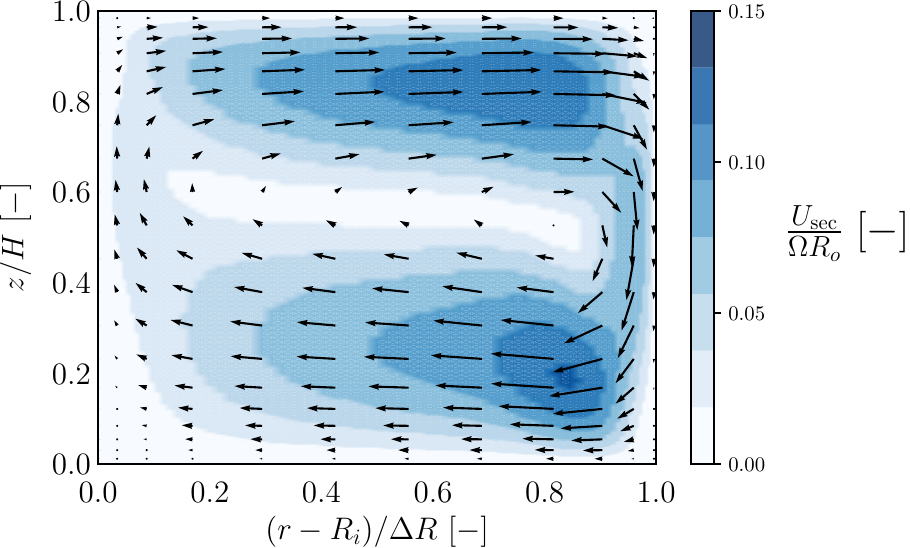}
\caption{\label{fig:snap5} Velocity field of the secondary flow in the radial-vertical plane for ${\rm Re} = 50$ ($A_c = 5$ and $C_r = 0.5$). The double arrow shows the characteristic height $H_{sec}$ of the secondary flow.}
\end{figure}

The first parameter of interest is the channel Reynolds number ${\rm Re}$. Its influence on the azimuthal velocity is studied in a case of moderate centrifugal and confinement effects, taking $C_r = \Delta R/R_0 = 0.5$ and $A_c = \Delta R/H=  5$. Fig.~\ref{fig:utz_re} shows the vertical profiles of the azimuthal velocity at various ${\rm Re}$. The case at the lowest Reynolds number (${\rm Re} = 5$) exhibits a linear profile similar to that of the SCPC flow, despite the presence of the wall confinement and centrifugal force. The influence of ${\rm Re}$, and hence of the centrifugal force (as discussed in Section \ref{sec:sec2}.B), becomes more pronounced at higher values, causing the profiles to converge towards a single S-shaped profile. This feature is likely to result from the reduction of the portion occupied by the secondary flow in the channel cross-section as ${\rm Re}$ increases, as it can be seen by comparing Fig.~\ref{fig:snap_used} and Fig.~\ref{fig:snap5}. When increasing ${\rm Re}$ from 50 to 280, the secondary flow compacts towards the walls while increasing in magnitude. A further increase of ${\rm Re}$ only amplifies the secondary-flow magnitude.
This leads to a higher alteration of the primary flow in these regions, ending in the S-shaped profile, which results from the redistribution of momentum by the secondary flow from the central region to the upper and lower regions.

The radial profiles, depicted in Fig.~\ref{fig:utr_re}, reveal a similar trend: as ${\rm Re}$ increases, they converge towards a given radial profile. At ${\rm Re} = 5$, the value of $U_\theta$ at middle radius is still equal to that of a SCPC flow, although a significant slope is present. Increasing ${\rm Re}$, the central part of the radial profile converges toward an asymptotic linear evolution. In parallel, we observe the development of a velocity peak in the outer region for the radial profiles (Fig.~\ref{fig:utr_re}), arising from the increasing magnitude of the secondary flow and its migration to the wall regions.

\subsubsection{Effect of the aspect and curvature ratios}

In order to investigate the influence of the channel aspect ratio $A_c$ and of the curvature ratio $C_r$, specific values have been selected to cover a wide range of these parameters: $C_r =(0.1, 0.5, 0.9)$, $A_c=(1, 2, 10)$, and ${\rm Re} =(5, 50, 500)$. Through $A_c$, the channel spans from a confined square section ($A_c = 1$) to an almost 2D channel ($A_c = 10$). The values of 0.1 and 0.9 of $C_r$ respectively corresponds to a weak and strong effect of the centrifugal force on the flow. The range of the channel Reynolds number covers the whole evolution of the flow structure in the laminar regime, starting from a linear profile (${\rm Re} = 5$), shifting to an intermediary curved profile (${\rm Re} = 50$) and ending with the sharp S-shaped profile (${\rm Re} = 500$).

\begin{figure*}[h!]
    \centering
        \includegraphics[scale=1.1]{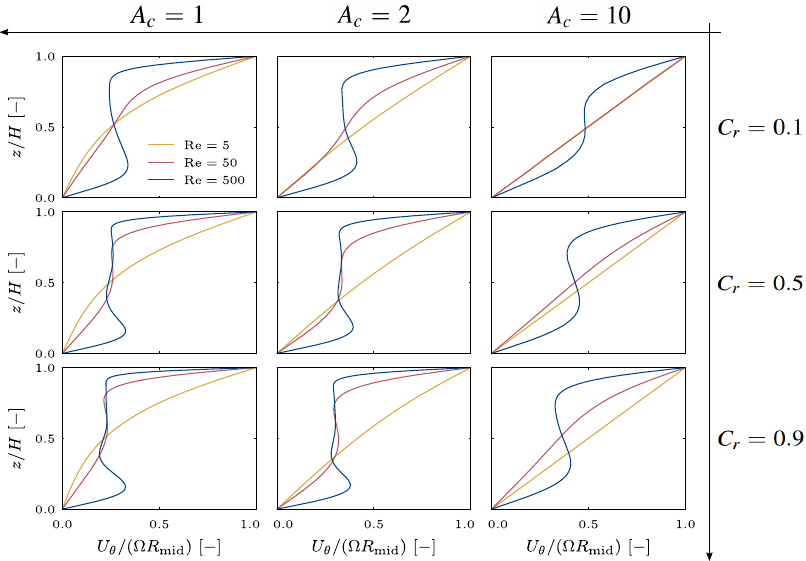}
    \caption{\label{fig:utz_map} Vertical profiles of the azimuthal velocity at $r=R_{\text{mid}}$ for different Reynolds numbers ${\rm Re} = (5,50,500)$. The different subplots correspond to their respective aspect ratio $A_c = \Delta R / H$ and curvature ratio $C_r = \Delta R / R_o$.}
\end{figure*}

Fig.~\ref{fig:utz_map} shows the evolution of the vertical profiles of the azimuthal velocity $U_{\theta}$ across the parameter map ($A_c$, $C_r$). For the lowest confinement and curvature ratios ($A_c=10$ and $C_r=0.1$, top right-hand corner of Fig.~\ref{fig:utz_map}), the velocity profile is linear for ${\rm Re} = 5$ and ${\rm Re} = 50$ (red and yellow curves), and is S-shaped at ${\rm Re} = 500$ (blue curve). As the curvature ratio $C_r$ increases (vertically downwards in Fig.~\ref{fig:utz_map}), the profiles tend to deviate from the linear trend, and the deviation is more pronounced the higher the Reynolds number. A similar effect is observed when increasing the degree of confinement (moving horizontally to the left).

\begin{figure*}[h!]
    \centering
        \includegraphics[scale=1.1]{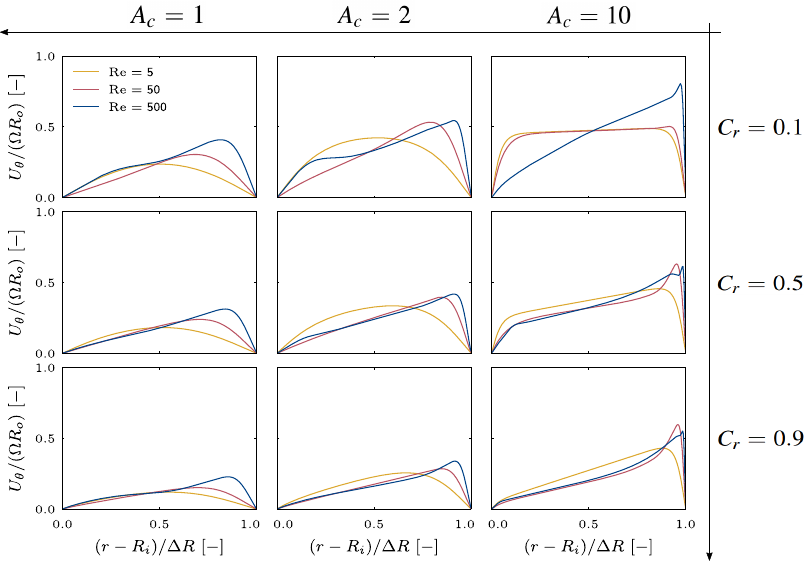}
    \caption{\label{fig:utr_map} Radial profiles of the azimuthal velocity at $z=Z_{\text{mid}}$ (middle height) for different Reynolds numbers ${\rm Re} = (5,50,500)$. The different subplots correspond to their respective aspect ratio $A_c = \Delta R / H$ and curvature ratio $C_r = \Delta R / R_o$.}
\end{figure*} 

As the confinement rate increases ($A_c$ decreases), the profiles deviate from the linear trend, and the deviation also increases with ${\rm Re}$. At the highest Reynolds number (${\rm Re}=500$), the S-curve tends to flatten in the upper central region as confinement increases ($A_c$ decreases), while gradients increase near the walls. Thus, for the plot at $A_c=1$ and $C_r=0.9$ (bottom left-hand corner of Fig.~\ref{fig:utz_map}), the profile at ${\rm Re} = 5$ is no longer linear but strongly curved, and the profile at ${\rm Re} = 500$ is flattened over 3/4 of the cross-section height. This flattening can be attributed to a significant increase of the influence of the secondary flow and of the confinement in these cases, which causes the increase of the velocity in the entire section. This results in a flat profile in the central to upper central region, reducing the boundary layer thickness close to the top moving wall. This effect is also observable in the lower central region, where the curve is steeper and the boundary layer thickness in the bottom region is decreased close to the bottom immobile wall.

Additionally, as $A_c$ or $C_r$ increases, the asymmetry of the velocity profile with respect to the half-height also increases. That behavior is reflected in the development of a near-wall gradient that is always stronger near the moving wall than near the immobile lower wall. The higher the curvature ratio $C_r$, the stronger this asymmetry is, which results from the secondary flow.

The radial profiles are reported in Fig.~\ref{fig:utr_map} and, as expected, a significant influence of $A_c$ and $C_r$ is also observed. For the case of lowest confinement and curvature effects ($A_c=10$ and $C_r=0.1$), the profiles are flat for ${\rm Re} = 5$ and ${\rm Re} = 50$ in the major part of the channel section, while the velocity increases almost linearly between the inner and the outer wall for ${\rm Re} = 500$. When the curvature ratio $C_r$ is increased, the profiles tend to adopt this linear trend at all Reynolds numbers. Increasing the confinement (i.e. decreasing $A_c$) turns out to reduce the linear part of the profile and curves the profiles, by increasing the gradient thickness near the wall in the radial direction.

In conclusion, the influence of the aspect ratio $A_c$ and the curvature ratio $C_r$ on the flow structure is important in an annular plane Couette flow. Up to moderate ${\rm Re}$, the vertical profiles are linear and the radial profiles are homogeneous in the channel section, provided that the confinement and curvature effects are weak (high aspect ratio $A_c$ and low curvature ratio $C_r$). In this configuration, at higher ${\rm Re}$, the $U_\theta$ profiles tends towards a S-shape in the vertical direction and a linear profile in the radial direction. Increasing the curvature ratio $C_r$ or decreasing the aspect ratio $A_c$ tends to distort the shape of the velocity profile, curving the linear profiles into a S-shape with a more and more flat central part and with steeper wall gradients. The value of ${\rm Re}$ at which this distortion appears is smaller as the effects of curvature and confinement become more pronounced. The selection of the APC configuration's geometrical parameters is therefore essential to ensure either a simpler 2D flow similar to SPC flow with a constant shear rate, or a more complex 3D flow. This choice will depend on the study and the intended applications.

\subsection{Experimental validation}\label{expsection}

\begin{figure}[h!]
     \centering
     \begin{subfigure}[h]{0.48\textwidth}
         \centering
         \includegraphics[width=0.9\textwidth]{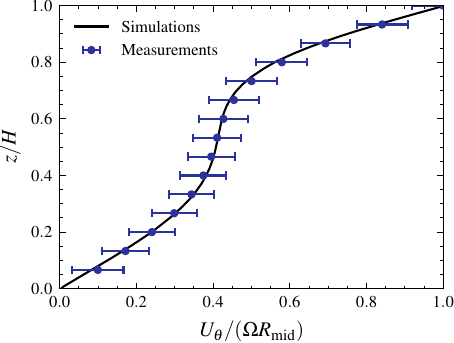}
         \caption{${\rm Re} = 300$}
         \label{fig:piv_300}
     \end{subfigure}
     \hfill
     \begin{subfigure}[h]{0.48\textwidth}
         \centering
         \includegraphics[width=0.9\textwidth]{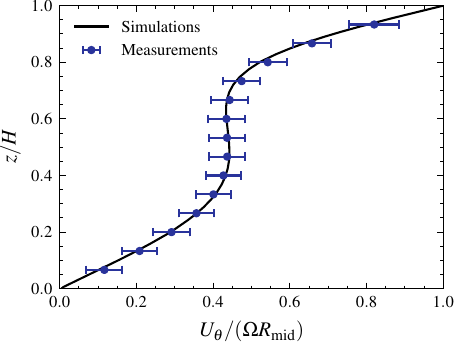}
         \caption{${\rm Re} = 400$}
         \label{fig:piv_400}
     \end{subfigure}
     \begin{subfigure}[h]{0.48\textwidth}
         \centering
         \includegraphics[width=0.9\textwidth]{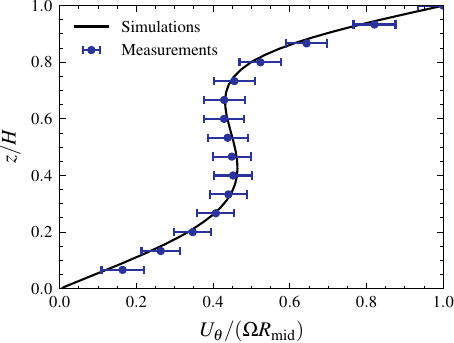}
         \caption{${\rm Re} = 500$}
         \label{fig:piv_500}
     \end{subfigure}
     \caption{Comparison of the vertical profiles of the azimuthal velocity obtained at $r=R_{\text{mid}}$ by PIV measurement and numerical simulations. The error bars on the measurements are quantified using the method from Wieneke \cite{WIENEKE2015}}
     \label{fig:piv}
\end{figure}

Using the experimental set-up described in Sec.~\ref{sec:expsetup}, the numerical simulations are tested against the PIV measurements of a low-viscosity oil flow in the APC channel. The vertical profiles of the azimuthal velocity $U_{\theta}$ at the middle radius $r = R_{\text{mid}}$ are obtained for three different Reynolds numbers: ${\rm Re} = (300,400,500)$ in order to address the S-shape profile regime. The results are plotted in Fig.~\ref{fig:piv}. The experimental measurements show only a slight deviation from the simulations, mostly in the top region. These results confirm the validity of the numerical simulations for the single-phase flow in the annular plane Couette channel.

\section{\label{sec:sec5} Scaling law for the torque and secondary flow}

In this section, we focus on the torque $\Gamma$ exerted by the fluid on the rotating top wall, which is a global quantity of primary interest. 
Firstly, it is directly related to the total viscous dissipation, $D=\Gamma\Omega$, and its scaling with the Reynolds number can provide insight into the flow dynamics.
Secondly, it can be used to determine the effective viscosity of a dispersed two-phase mixture, which is an important issue for predicting the dynamics of suspensions of solid particles \cite{GUAZZELLI2018} as well as droplet emulsions \cite{YI2021, Yi2022, ABBAS2017}, especially at large Reynolds numbers where experimental results are rare. The detailed description of the velocity field in the APC flow for a wide range of parameters showed a rather complex picture. Thus, obtaining simple laws for $\Gamma$ might seem unlikely. However, this was achieved for a single-phase Taylor Couette (TC) flow at large-Reynolds-number  by Eckhardt et al. \cite{ECKHARDT2000, ECKHARDT2007} by drawing an analogy between momentum transfer in this flow and the heat transfer by convection in a Rayleigh-B\'enard flow. Later, this approach was successfully applied to the experimental determination of the effective viscosity of an emulsion in a TC flow by Yi et al. \cite{YI2021,Yi2022}, who adjusted the parameters of the scaling law to the multiphase case. In the following, we adapt the method developed by those authors to find scaling laws for the APC flow. Before doing so, it is worth mentioning the differences and similarities between the TC and the APC configurations. In both cases, the velocity gradients concentrate towards the walls as the Reynolds number increases. However, the transition from the linear solution at small Reynolds number to the nonlinear solution at large Reynolds number is not of the same nature. In the laminar regime for the TC case, only the azimuthal velocity is not zero and the non-linear terms in the azimuthal momentum equation are identically null by symmetry. Beyond a critical Reynolds number, the steady laminar solution becomes unstable and the scaling law found by 
 \cite{ECKHARDT2000, ECKHARDT2007, Yi2022, YI2021} was obtained in the turbulent regime. In the APC case, the secondary flow in the radial and vertical directions is already present at any Reynolds number and increases regularly with ${\rm Re}$ until it plays a major role on the shape of the primary flow. Even though the present results show a S-shape profile similar to that observed in a turbulent APC flow, they are obtained in the laminar regime.

\subsection{Case of the Taylor-Couette flow}

We begin by recalling the approach introduced for the TC flow by \cite{ECKHARDT2000, ECKHARDT2007, Yi2022, YI2021}.
Assuming a steady axisymmetric flow, an analytical solution can be derived in the laminar regime. When the inner cylinder rotates at a constant angular velocity $\omega_i$ while the outer cylinder is immobile, the torque is given by
\begin{equation}
    \Gamma_{\text{TC,0}} = 4 \pi \mu r_i^2 H\dfrac{\omega_i}{1 - (r_i/r_o)^2}\,,
\end{equation}
\noindent where $r_i$ and $r_o$ are the inner and outer cylinder radii, $H$ is the cylinder height and $\mu$ the dynamic viscosity of the pure fluid \cite{ARIAS2015}.
In order to find a scaling law for large Reynolds numbers, it is relevant to rewrite this expression by making the role of the Reynolds number explicit.
That can be achieved by introducing a viscous torque scale that is independent of the angular velocity \cite{ECKHARDT2000, DUBRULLE2000},
\begin{equation}
    \Gamma_{\text{TC},\mu} = \dfrac{2 \pi \mu^2 H}{\rho}\,.
\end{equation}
Normalizing the laminar torque by it, it yields
\begin{equation}
    \Gamma^*_{\text{TC,0}} = \dfrac{\Gamma_{\text{TC,0}}}{\Gamma_{\text{TC},\mu}} = K_0{\rm Re}_{\text{TC}},
\label{eq:TC0}
\end{equation}
where the Taylor Couette Reynolds number writes 
\begin{equation}
    \rm{Re}_{TC} = \frac{\rho \omega_i r_i (r_o-r_i)}{\mu},
\end{equation}
and $K_0$ depends only on the geometry,
\begin{equation}
    K_0 = \frac{2(r_i/r_o)}{(1+r_i/r_o)(1-r_i/r_o)^2}\,.
\end{equation}
Beyond the laminar regime, the linear evolution of the torque with the Reynolds number is no longer valid. Eckhardt et al. \cite{ECKHARDT2000, ECKHARDT2007} proposed to generalize Eq.~\ref{eq:TC0} to high Reynolds number turbulent flows as 
\begin{equation}
    \Gamma^* = \dfrac{\Gamma}{\Gamma_{\text{TC,}\mu}} = K_{\infty} {\rm Re}_{\text{TC}}^{\alpha}\,,
\label{eq:GTC}
\end{equation}
where the prefactor $K_{\infty}$ only depends on the geometry.
Various works found the power-law exponent $\alpha$ to range from $1.5$ to $1.7$ in a single-phase flow \cite{ECKHARDT2000,DUBRULLE2000,ARIAS2015,Yi2022, YI2021}.

Yi et al. \cite{YI2021,Yi2022} applied the same approach to the flow of a droplet emulsion by assuming that the two-phase mixture can be described as an equivalent homogeneous fluid of effective density $\rho_m$ and viscosity $\mu_m$, which are used in the definitions of $\Gamma_{\text{TC},\mu}$ and ${\rm Re}_{TC}$ instead of the pure fluid values. The experimental measurements of the torque led to $\alpha=1.58$, a value similar to that of a single-phase flow. Thus, assuming $K_{\infty}$ in Eq.~\ref{eq:GTC} are the same in single-phase and two-phase flows, an effective viscosity of the emulsion was determined.

\subsection{Case of the APC flow}

\begin{figure}[h!]
\centering
\includegraphics[width=0.7\textwidth]{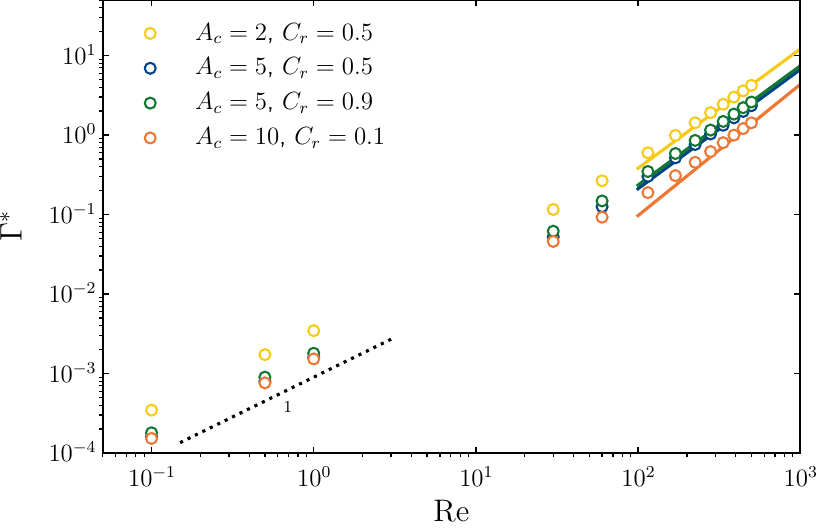}
\caption{\label{fig:torque} Dimensionless torque $\Gamma^* = \Gamma/\Gamma_{{\rm APC},\mu} $ versus the Reynolds number ${\rm Re}$ for various pairs of geometrical parameters: ($A_c=2$, $C_r = 0.5$), ($A_c=5$, $C_r = 0.5$), ($A_c=5$, $C_r = 0.9$) and ($A_c=10$, $C_r = 0.1$). Symbols: numerical simulations; straight lines: high-${\rm Re}$ fits by $A_{\infty}\, {\rm Re}^{\alpha}$. The fitted parameters are respectively $A_0 = (3.5\times 10^{-3}, 1.6\times 10^{-3}, 1.8\times 10^{-3}, 1.5\times 10^{-3})$, $A_{\infty} = (3.8\times 10^{-4}, 2.1\times10^{-4}, 2.3\times10^{-4}, 0.51\times10^{-4})$ and $\alpha = (1.5, 1.5, 1.5, 1.65)$.}
\end{figure}

We now analyze the torque in the annular plane Couette configuration. As opposed to the TC case, because the azimuthal velocity at the top moving wall varies with the radial position ($U_\theta=\Omega$r), the velocity field at any finite Reynolds number is three-dimensional and does not admit a trivial analytical solution. Instead, to build a reference torque $\Gamma_0$, we consider the shear stress $\tau_{\theta z} = \mu U/H$ from the Straight Plane Couette flow in which we substitute the velocity $U$ by $\Omega R_o $. Then, by integrating this stress over the surface $S_{top} = \pi (R_o^2 - R_i^2)$ of the top wall and by using $R_o$ as the lever arm, it yields  
\begin{equation}
\label{eq:T_st_apc}
    \Gamma_0  = \mu \dfrac{\Omega R_o}{H} \pi (R_o^2 - R_i^2)R_o\,.
\end{equation}
Dividing by the APC Reynolds number $\rm{Re}$, we get the following viscous torque scale
\begin{equation}\label{Gamu}
    \Gamma_{\mu} = \dfrac{\Gamma_0}{{\rm Re}} = \dfrac{4 \pi \mu^2 (R_o^2 - R_i^2)R_0}{ \rho H^2}.
\end{equation}
The torque $\Gamma$ exerted by the flow on the top annular plate can be computed from the numerical results by integrating the local wall shear stress ($\mu \partial U_{\theta}/\partial z$) multiplied by the radial coordinate over the surface of the top plate. Following the approach developed for the TC flow \cite{ECKHARDT2000,DUBRULLE2000}, we can expect the dimensionless torque $\Gamma^*$ to behaves as 
\begin{equation}
\label{eq:G_apclowRe}
    \Gamma^* = \frac{\Gamma}{\Gamma_{\mu}}=A_0 {\rm Re}\,,
\end{equation}
at low ${\rm Re}$ and as
\begin{equation}
\label{eq:G_apclargeRe}
    \Gamma^* = \frac{\Gamma}{\Gamma_{\mu}}=A_{\infty} {\rm Re}^{\alpha}\,,
\end{equation}
at large ${\rm Re}$.
Figure~\ref{fig:torque} presents $\Gamma^*$ as a function of the Reynolds number ${\rm Re}$ for various geometries defined by different parameter pairs of $(A_c,C_r)$. The symbols show the torque computed from the numerical results and the plain straight lines its corresponding fit by Eq.~\ref{eq:G_apclargeRe} for each geometry. The values of $\alpha$, $A_0$ and $A_{\infty}$ are given in the caption.

At low to moderate ${\rm Re}$, the linear regime is clearly present in all cases. The value of $A_0$ is independent of $C_r$, since $C_r$ is associated to the centrifugal effect which is negligible in this regime. Moreover, it also becomes independent of the lateral confinement when $A_c$ becomes large. The transition to the large-${\rm Re}$ asymptotic regime takes place in the range from ${\rm Re}=10$ to 100, starting and ending sooner as $C_r$ increases. At large ${\rm Re}$, $ \Gamma^*$ is well described by a power law, $A_{\infty}\, {\rm Re}^{\alpha}$, with an exponent in the same range as in a TC flow: $\alpha=1.5$ for $C_r \ge 0.5$ and $\alpha=1.65$ for $C_r \le 0.1$ while $A_{\infty}$ depends on both $A_c$ and $C_r$. It is out of the scope of this work to perform an exhaustive investigation of the effect of the geometry. However, the present results indicate that, in cases of low confinement ($A_c\ge5$), the value of $A_{\infty}$ increases with $C_r$, as the centrifugal effect is enhanced. 

\begin{figure}[h!]
\centering
\includegraphics[width=0.7\textwidth]{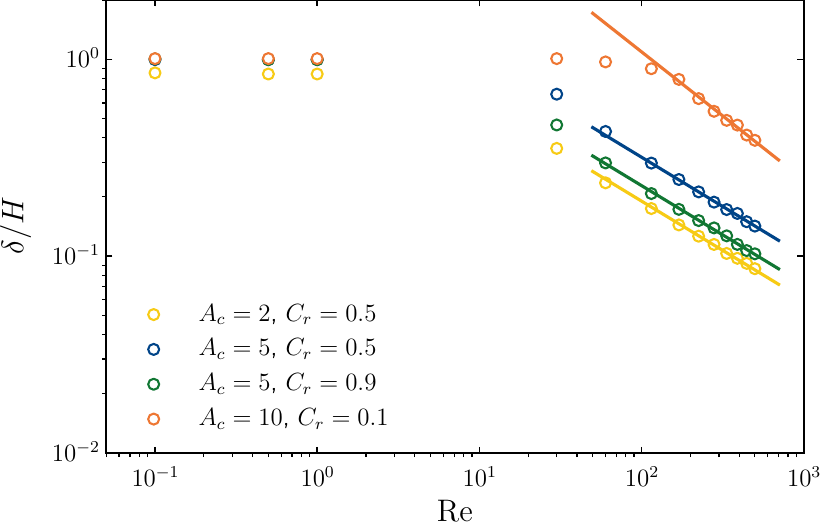}
\caption{\label{fig:delta} Dimensionless boundary-layer thickness of the primary flow at the top wall.
Symbols: numerical simulations for the same pairs of parameters as in Fig.~\ref{fig:torque}; straight lines: high-${\rm Re}$ fits by $B_{\infty}\, {\rm Re}^{1-\alpha}$ with $B_{\infty}=(1.9, 3.18, 2.3, 21.8)$ and $\alpha = (1.5, 1.5, 1.5, 1.65)$.}
\end{figure}

Increasing the Reynolds number, the velocity gradients concentrate near the walls. It is thus interesting to examine the evolution of the boundary layer thickness $\delta$ of the primary flow. Here it is estimated from the local shear strain ($\partial U_\theta/\partial {z}$) on the top moving wall ($z=H$) at the middle of the channel ($r=R_{mid}$), i.e. it is estimated as $(\partial U_\theta/\partial {z}) \delta=\Omega R_{mid}$. Figure~\ref{fig:delta} shows the evolution of $\delta/H$ against ${\rm Re}$ for the same cases as in Fig.~\ref{fig:torque}. Note that if the shear stress was uniform over the entire top wall, there should be a direct relation between $\delta/H$ and $\Gamma^*$,  an evolution of $\Gamma^*$ as ${\rm Re}^\alpha$ involving an evolution of $\delta/H$ as ${\rm Re}^{1-\alpha}$. The numerical results show that this relation is satisfied in the linear regime as well as in the large-${\rm Re}$ regime. At low ${\rm Re}$, $\delta/H$ is almost unity, as expected for a constant shear rate along the channel height. At large ${\rm Re}$, $\delta/H$ indeed scales as ${\rm Re}^{1-\alpha}$, with the same value of $\alpha$ as that obtained from the fit of $\Gamma^*$. The fact that a local measurement is sufficient to characterize the global feature of the flow has major consequences. Regarding physical mechanisms, it suggests that the entire flow reaches a self-similar asymptotic state.

Practically, it shows that the evolution of the torque and total dissipation can be obtained from the measurement of the velocity profile of the primary flow in a single plane, such as those presented in section~\ref{expsection}.

\begin{figure}[h!]
\centering
\includegraphics[width=0.7\textwidth]{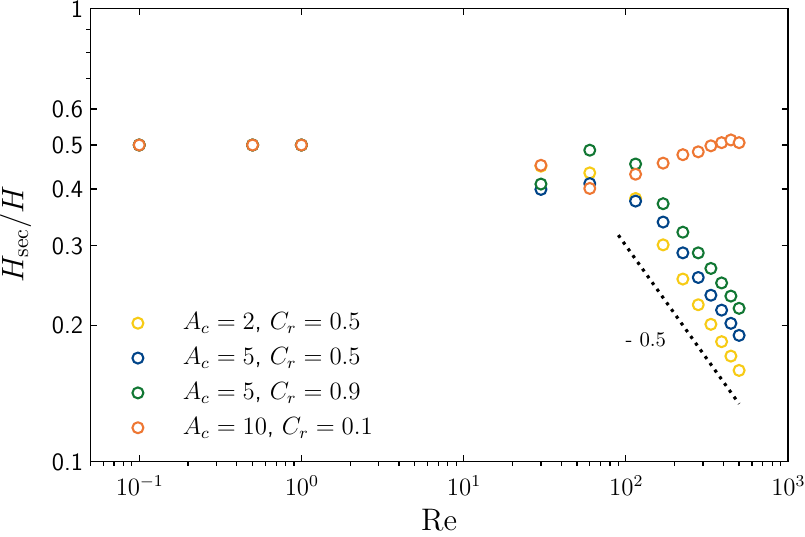}
\caption{Dimensionless characteristic height of the secondary flow for the same pairs of parameters as in Fig.~\ref{fig:torque}.}
\label{fig:Hsed}
\end{figure}

\begin{figure}[h!]
\centering
\includegraphics[width=0.7\textwidth]{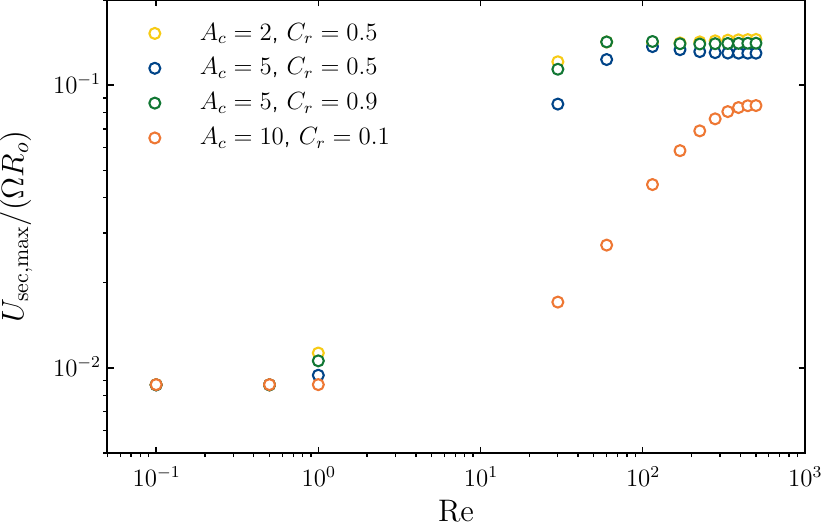}
\caption{Dimensionless characteristic velocity of the secondary flow for the same pairs of parameters as in  Fig.~\ref{fig:torque}.}
\label{fig:Used_max}
\end{figure}

We examine now the scaling of the secondary flow in the two asymptotic regimes. The secondary flow in the radial-vertical plane consists of a single recirculation cell (Fig.~\ref{fig:snap5}). In order to characterize it, we introduce two parameters: its maximal magnitude $| U_{sec,max} |$ and the vertical distance $H_{sec}$ between the two points located in the lower half of the domain where the secondary-flow magnitude is $0.3 | U_{sec,max} |$ (see Fig.~\ref{fig:snap5}). Figures~\ref{fig:Hsed} and \ref{fig:Used_max} present the evolution of $H_{sec}$ and $U_{sec,max}$ against ${\rm Re}$. At low ${\rm Re}$,  all cases show a similar trend: $H_{sec}=\frac{1}{2} H$, which means that the secondary-flow cell occupies the entire cross section, and the velocity $U_{sec,max}$ of the secondary flow is two orders of magnitude smaller than that of the primary flow. At large ${\rm Re}$, the three cases with a significant curvature ratio ($C_r\ge 0.5$) behave in the same way. As ${\rm Re}$ increases,
the secondary-flow cell localizes more and more near the walls, $H_{sec}$ showing the same scaling law in ${\rm Re}^{1-\alpha}$ as $\delta$. In parallel, the velocity of the secondary flow reaches a plateau ($0.13\le U_{sec,max}/\Omega R_o \le 0.15$) that slightly depends on the geometry. Thus, the shear rate characteristic of the primary flow ($\Omega R_o/\delta$) and that of the secondary flow ($U_{sec,max}/H_{sec}$) obey the same scaling law with ${\rm Re}$, which confirms the close connection between primary and secondary flows in this regime. In the case of a small curvature ratio ($C_r\ge 0.1$), centrifugal inertia is reduced and the transition towards the large-${\rm Re}$ asymptotic regime is delayed. The plateau of $U_{sec,max}$ is hardly reached at the largest investigated ${\rm Re}$, while the power-law decrease of $H_{sec}$ is not attained at all.

\section{\label{sec:sec6}Conclusion}

The goal of this work was to understand the hydrodynamics of the flow within an Annular Plane Couette (APC),  a peculiar configuration involving secondary flows of Prandtl's first kind that can be used as a reference flow for fundamental studies, such as investigations of the rheology of multiphase media. Direct numerical simulations of a single-phase, laminar, stationary, axisymmetric, incompressible, Newtonian flow within an APC channel have been carried out using \textit{OpenFOAM}. These simulations were further validated by PIV measurements in an experimental APC device at various Reynolds numbers, using a low-viscosity oil seeded by fluorescent particles.

The numerical study focuses on analyzing the effects of three dimensionless parameters: (1) the channel Reynolds number ${\rm Re}$, which compares inertial to viscous forces, (2) the channel aspect ratio $A_c$, which characterizes the lateral wall confinement, and (3) the curvature ratio $C_r$, which modulates the influence of the centrifugal forces. Initially, the main features of the hydrodynamics are examined in a single representative case, then the APC flow is scanned by varying the geometric parameters within broad ranges.

The fluid is sheared due to the rotation of the top annular plate, inducing a primary flow in the azimuthal direction, akin to the one observed in a Straight Plane Couette (SPC) flow. However, the vertical profile of the azimuthal velocity deviates from the classical linear profile, showing a S-shaped form reminiscent of a turbulent SPC flow. A secondary flow, in the form of a single recirculation cell in the radial-vertical plane, is also generated by the centrifugal force difference across the radial direction. This secondary flow has a magnitude up to 15\% of the primary flow. It thus exerts a significant influence on the primary flow by increasing the shear rate close to the walls and decreasing it in the central region.

Regarding the parametric study, several key observations can be drawn. The results show that as ${\rm Re}$ increases, the flow evolves from a linear profile to a S-shape profile, due to the increased localization and magnitude of the secondary flow in the near-wall regions. The effects of lateral confinement are significant for a high confinement ($A_c = 1$), but they notably diminish once the section becomes wider ($A_c \geq 2$), and become negligible when the section is almost 2D ($A_c = 10$). Meanwhile, increasing the curvature ratio $C_r$ leads to a transformation of the vertical profile of the azimuthal velocity into a S-shape, while lowering it results in an almost 2D flow, fairly similar to the SPC flow. Especially, for the case of low confinement and centrifugal forces ($A_c=10$ and $C_r=0.1$), the linear profile is maintained at higher Reynolds numbers, while the velocity profile is kept uniform in the radial direction.

Then, the torque exerted on the top moving wall has been examined. The dimensionless torque $\Gamma^*$ is a global quantity that characterizes the flow as a whole. Its evolution with the Reynolds number clearly exhibits the existence of two asymptotic regimes. At low ${\rm Re}$, $\Gamma^*$ is a linear function of ${\rm Re}$, with a slope $A_0$ depending on the lateral confinement through the value of $A_c$. At large ${\rm Re}$, it is a power law, $\Gamma^*=A_{\infty} {\rm Re}^{\alpha}$, with an exponent $\alpha$ close to 1.5, a value in the same range as those measured in a turbulent Taylor Couette flow of either a single fluid \cite{ECKHARDT2000, ECKHARDT2007} or a droplet emulsion \cite{Yi2022, YI2021}. Such similarities, between a laminar APC flow and either the turbulent TC flow regarding the torque scaling or the turbulent SPC flow regarding the S-shaped velocity profile, suggest an analogy of the underlying mechanisms. As ${\rm Re}$ increases, our results show that the velocity gradients concentrate near the wall, in a boundary layer whose thickness $\delta$ scales as ${\rm Re} ^{1-\alpha}$. An important result is that various global or local quantities of either the primary or the secondary flow show all the same scaling with the Reynolds number. This indicates that, for a given geometry, the structure of the APC flow becomes self-similar at large Reynolds numbers. Practically, it means that a local measurement, such as the velocity gradient at the top wall in the channel middle can be enough to characterize the evolution of the torque, as well as the total viscous dissipation by the flow.

In conclusion, the annular plane Couette flow can be used as a reference flow despite the centrifugal effects due to curvature. Indeed, by enabling long-duration studies, a shear plane parallel to gravity and a controllable shear rate, this configuration seems to open up new possibilities for the investigation of more complex processes, especially in emulsion flows, with slow evolution of the interfacial area. By adjusting the Reynolds number (${\rm Re}$) and the geometry ($A_c$ and $C_r$), the flow field within the channel can extend from an almost 2D flow, close to the classical plane Couette flow and its constant shear rate, to a complex 3D flow, more representative of industrial applications where the shear rate in pipe flows is not uniform.

\begin{acknowledgments}
The authors would like to thank the technical department of the LGC and of IMFT for their help to construct the experimental set-up, Thomas Ménager for the construction and tuning of the experimental set-up and Emmanuel Cid for his help on the development of the optical metrology.
\end{acknowledgments}

\appendix

\bibliography{ref}

\end{document}